\documentclass[aps,prc,twocolumn,showpacs,superscriptaddress]{revtex4}
\usepackage[latin1]{inputenc}
\usepackage{bm}
\usepackage{epsfig}
\usepackage{amsthm}
\usepackage{amsfonts}
\usepackage{float}
\usepackage{amsmath,amssymb}
\usepackage{graphicx}
\usepackage{dcolumn}
\usepackage{bm}
\usepackage[colorlinks = true,
            linkcolor = red,
            urlcolor  = red,
            citecolor = blue,
            anchorcolor = green]{hyperref}
\newcommand{\be}{\begin{equation}}
\newcommand{\ee}{\end{equation}}
\newcommand{\bea}{\begin{eqnarray}}
\newcommand{\eea}{\end{eqnarray}}

\newcommand{\del}{\partial}

\begin{document}
\title{Effects of causality on the fluidity and viscous horizon of quark-gluon plasma} 
\author{Mahfuzur Rahaman}
\email{mahfuzurrahaman01@gmail.com}
\affiliation{Variable Energy Cyclotron Centre, 1/AF, Bidhan Nagar, Kolkata - 700064, India}
\affiliation{Homi Bhabha National Institute, Training School Complex, Anushaktinagar, Mumbai - 400085, India}
\author{Jan-e Alam}
\email{jane@vecc.gov.in}
\affiliation{Variable Energy Cyclotron Centre, 1/AF, Bidhan Nagar, Kolkata - 700064, India}
\affiliation{Homi Bhabha National Institute, Training School Complex, Anushaktinagar, Mumbai - 400085, India}

  
\begin{abstract}
The second order Israel-Stewart-M$\ddot{u}$ller relativistic hydrodynamics has been applied to study the effects of causality 
on the acoustic oscillation in relativistic fluid.  Causal dispersion relations have been derived with non-vanishing 
shear viscosity, bulk viscosity and thermal conductivity at non-zero temperature and baryonic chemical potential. 
These relations  have been used to investigate the fluidity of Quark Gluon Plasma (QGP) at finite temperature ($T$).
Results of the first order dissipative hydrodynamics have been obtained as 
limiting case of the second order theory. The effects of the causality on the fluidity near the transition 
point and on viscous horizon are found to be significant. We observe that the inclusion of causality increases the 
value of fluidity measure of QGP near $T_c$ and hence makes the flow strenuous.  
It has also been shown that the inclusion of large magnetic field in the causal hydrodynamics alters 
the  fluidity of QGP. 
\end{abstract}
\maketitle
%
%
%
\section{Introduction}
The collision of heavy ions at relativistic energies create matter in a new 
state called quark-gluon plasma (QGP)~\cite{rhic1,lhc1}. The QGP can be created 
with different temperatures ($T$) and net baryonic chemical potential ($\mu$) 
by altering the energy of the colliding beams~\cite{cbmbook}. 
For example, the system formed in the nuclear collisions at Large Hadron 
Collider (LHC) as well as at highest the Relativistic Heavy Ion Collider (RHIC) energy
will have very small
$\mu$ but large $T$. On the other hand the matter created at GSI-FAIR (Facility for Antiproton and Ion Research) 
energy, JINR-NICA (Nuclotron-based Ion Collider fAcility) and at
lower energy run of RHIC will  have larger
$\mu$ but smaller $T$.
Nature of the transition from QGP to hadrons depends on the values of $T$ and
$\mu$~\cite{fodor}.  It is expected that at high $\mu$ and low $T$ the phase
transion is first order but at high $T$ low $\mu$ it is
a continuous transition from QGP to hadrons~\cite{lattice1,lattice2,Katz,Karsch_lat}.
When the QGP reverts to hot hadrons due to cooling caused by expansion,
the system may encounter the critical point in the QCD phase
diagram during the transition from QGP to hadrons. 
The characterization of the fluid at the critical point is one of the most
crucial problem in  heavy ion collision at relativistic energies.

Lattice QCD simulations at zero $\mu$ indicate that strongly interacting 
nuclear matter undergoes a rapid transition from a chirally broken confined hadronic phase 
to a chirally symmetric, deconfined QGP 
around $T_c\sim $155 MeV~\cite{Karsch_lat}. The QGP
expands very fast  due to internal pressure and its evolution  
in  space-time can be studied by using relativistic viscous hydrodynamics. 
In general, the presence of non-zero
transport coefficient, like shear and bulk viscosities and 
thermal conductivity make the evolution and characterization of QGP very challenging and complex.  
The  Navier-Stokes equation is not suitable to describe
relativistic fluid  as it suffers 
from severe flaws, {\it e.g.} it violates  causality and leads to unstable solutions~\cite{Hiscock}.
These unphysical behaviors were resolved by M$\ddot{u}$ller\cite{Muller} using Grad's 
14 moment method\cite{Grad} and its relativistic covariant form is due to Israel and 
Stewart\cite{Israel79}. These  theories are based on extended irreversible thermodynamics
known as second order theories. 
The first order and second order hydrodynamical descriptions stem from the definition 
of entropy four-current. 
The conservations of energy-momentum and conserved charge (e.g. net baryon number) 
along with the second law of 
thermodynamics lead to the dynamical transport equations which are hyperbolic in nature 
and respect causality.
 
The transport coefficients such as shear viscosity, bulk viscosity, thermal conductivity
etc. are taken as input in 1st order  hydrodynamics. 
In addition to these standard transport 
coefficients, the causal or 2nd order theory contains a few  more thermodynamic functions which are known 
as second-order coefficients. These coefficients along with the standard transport 
coefficients, correspond to different relaxation times and relaxation lengths for various 
dissipative fluxes which are absent in acausal theory. 
The results  of acausal theory can be obtained 
by setting these extra coefficients to  zero in causal theory. 
In this work we use the relativistic causal hydrodynamics to investigate 
propagation of acoustic wave through dissipative fluid 
with non-zero net (baryonic) charge, shear viscosity,
bulk viscosity and thermal conductivity following  the
procedure outlined in Ref.\cite{Weinberg71}. 
In the present work we investigate the effects of causality 
on the fluidity of QGP in contrast to earlier
work where the fluidity of QGP has been studied ~\cite{koch_prc} within the scope of 
first order theory which is flawed due to causality violation in the relativistic domain.  
The aim of this work is to estimate the shift on the fluidity of relativistic
fluid by using second order hydrodynamics which respects causality.
Maartens et al.~\cite{Maartens}
has used causal hydrodynamics to  explore the dissipation of acoustic waves in
baryon-photon fluid in early universe.

The present article is organized as follows: In Section \ref{sec:form}, we will discuss  the formalism used to derive
the transverse and longitudinal dispersion relations for sound wave 
within the framework of causal hydrodynamics. Dispersion relations for sound wave in the
dissipative system with the inclusion of  magnetic field have been derived in section \ref{sec:3}. 
The impact of the causality and external magnetic field on the fluidity  
has been discussed in section \ref{sec:4}. Section \ref{sec:5} has been devoted to present results
and finally section \ref{sec:concl} has been dedicated to summary and discussions.
We have used natural unit,  {\it i.e.} $c=\hbar=k_B=1$  here
and the  Minkowski metric is set as  $g^{\lambda\mu}=diag(-,+,+,+)$.

\section{Formalism: derivation of causal dispersion relations}
\label{sec:form}
The  relativistic energy-momentum tensor ($T^{\lambda\mu}$) in the  Israel-Stewart\cite{Israel79}
second order theory is given by:
\begin{equation}
T^{\lambda\mu}=\epsilon u^\lambda u^\mu+P\triangle^{\lambda\mu}+2 h^{(\lambda} u^{\mu)}+\tau^{\lambda\mu}
\label{eq1}
\end{equation} 
where the dissipative viscous stress tensor $\tau^{\lambda\mu}=\Pi\triangle^{\lambda\mu}+\pi ^{\lambda\mu}$ 
with $\pi^\lambda_\lambda=h^\lambda u_\lambda=\tau^{\lambda\mu}u_\lambda =0$ 
where the projection operator is defined by $\triangle^{\lambda\mu}=g^{\lambda\mu}
+u^{ \mu} u^{\lambda }$ with $ u^{ \mu} u_{\mu }=-1$.   
The heat flux four vector is given by $q^\mu =h^\mu-n^\mu (\epsilon+P)/n$,  
the particle four flow 
$N^\mu =nu^\mu +n^\mu $ with $n^\mu u_\mu =0$,   where 
$n$ is the net number density, $\Pi $ is the bulk pressure,        
$u^\mu $ is the fluid four velocity,   $\epsilon$ is the energy density,    
$P$ is the thermodynamic pressure and $h(=\epsilon+P)$ is the enthalpy density. The  symmetric tensor $h^{(\lambda} u^{\mu)}$ is defined as 
$h^{(\lambda} u^{\mu)}=\frac{1}{2}(h^{ \lambda} u^{\mu }+ h^{ \mu} u^{\lambda })$. 

The definition of fluid four velocity in Eq.\ref{eq1} can be fixed 
by choosing a proper reference frame 
attached to the fluid element either due to Landau-Lifshitz (LL)  or 
Eckart.  
The Eckart frame~\cite{Eckart} represents a local rest frame for which the net charge dissipation is zero
but the net energy dissipation is non-zero.
The LL frame~\cite{Landau} represents a local rest frame where
the energy dissipation is zero but the net charge dissipation is non-zero.
We consider LL frame here to study a system with net non-zero charge (baryon number).

In LL frame: $h^\mu =0$, $n^\mu =-n q^\mu /(\epsilon+P)  $ and the different viscous fluxes are given by \cite{Israel79}
\begin{equation}\begin{aligned}
 & \Pi =-\frac{1}{3}\zeta(u^\mu _{|\mu }+\beta_0 D \Pi-\alpha_0  q^\mu _{|\mu })\\& q^\lambda=\chi T\triangle^{\lambda\mu}[(\del_\mu \alpha )nT/(\epsilon+P)-\beta_1 D{q_\mu}+\alpha_0\del_\mu \Pi +\alpha_1\Pi ^\nu _{\mu|\nu}]\\&\Pi_{\lambda\mu }=-2\eta[u_{<\lambda |\mu >}+\beta_2D{\Pi} _{\lambda\mu }-\alpha_1q_{<\lambda |\mu >}]
 \end{aligned}
 \label{eq2}
\end{equation}
where  $D\equiv u^\mu\del_\mu$, is  well known 
 co-moving derivative or material derivative. In the local rest frame, $D\Pi=\del_0\Pi\equiv\dot{\Pi}.$ 
The different coefficients appearing in Eq.\ref{eq2} are  $\alpha=\mu/T$, is known as thermal potential, 
 $\zeta $ is the coefficient of bulk viscosity,
 $\eta$ is the coefficient of shear  viscosity, 
 $\chi$ is the coefficient of thermal conductivity,  
 $\beta _0,\beta_1,\beta_2$ are relaxation coefficients, 
 $\alpha_0$ and $\alpha_1$ are coupling coefficients. The relaxation times for the 
bulk pressure ($\tau_{\Pi}$), the heat flux ($\tau_q$) and the shear tensor ($\tau_{\pi}$) 
are defined as~\cite{muronga} 
\begin{equation}
\tau_{\Pi}=\zeta \beta_0, \tau_q=k_BT\beta_1, \tau _{\pi}=2\eta \beta_2
\label{eq3}
\end{equation}
The relaxation lengths which  couple the heat flux and bulk  pressure 
($l_{\Pi q}, l_{q\Pi} $), the  heat flux and shear tensor $(l_{q\pi}, l_{\pi q})$ are defined as 
\begin{equation}
l_{\Pi q}=\zeta \alpha_0, l_{q\Pi} =k_B T \alpha _0, l_{q\pi}=k_BT\alpha_1, l_{\pi q}=2\eta \alpha_1 
\label{eq4} 
\end{equation}  
The symmetric, trace free part of the spatial projection is defined by 
$A_{<\lambda \mu >}\equiv[\triangle^\alpha _{(\lambda}\triangle^\beta _{\mu)}-\frac{1}{3}\triangle_{\lambda\mu }
\triangle^{\alpha \beta}]A_{\alpha \beta}$ and $u^\mu _{|\mu }\equiv \del_\mu u^\mu $.
Since in energy frame  $h^\mu=0$, then the energy-momentum tensor reduces to
\begin{equation}
T^{\lambda\mu}=\epsilon u^\lambda u^\mu+P\triangle^{\lambda\mu} +\Pi\triangle^{\lambda\mu}+\pi ^{\lambda\mu} 
\label{eq5}
\end{equation}
We put the explicit forms of $\Pi, q^\lambda$ and $\pi ^{\lambda\mu}$ given by Eq.\ref{eq2} into Eq.\ref{eq1} to get 
\begin{equation}
\begin{aligned}
T^{\lambda\mu}&=  \epsilon u^\lambda u^\mu+P\triangle^{\lambda\mu}-
\frac{1}{3}\zeta u_{|\sigma} ^\sigma \triangle^{\lambda\mu}+
\frac{1}{9} \zeta^2\beta_0 \dot{u}_{|\rho }^\rho   \triangle^ {\lambda\mu}\\& + 
\frac{nT^2}{P+\epsilon}\frac{\zeta \alpha _0\chi  }{3}\del_\sigma [\triangle ^{\sigma\rho }
(\del_\rho \alpha)\triangle^{\lambda\mu}]-
2\eta u^{<\lambda|\mu >}\\& + 4\eta^2\beta_2 \dot{ u}^{<\lambda|\mu >}+\frac{2nT^2}{P+\epsilon}\alpha _1\eta\chi [\triangle^{(\lambda} _\alpha\triangle^{\mu)} _{\beta}\\&-\frac{1}{3}\triangle_{\alpha \beta}
\triangle^{\lambda\mu }] \del^{\beta} \triangle^{\alpha \rho }\del_\rho \alpha 
\end{aligned}
\label{eq6}
\end{equation}
where we have kept terms upto  second order in space-time derivatives and neglected all the higher order space-time derivatives.
We impart small perturbations $P_1,\epsilon_1, n_1, T_1$ and $ u^\alpha_1$  to
$P, \epsilon, n, T$ and $u^\alpha $  respectively to study 
the acoustic oscillations set by these perturbations.  In this work we
consider a non-expanding fluid with $u^\alpha=(1,0,0,0)$.  Then the 
perturbation, $u_1^\alpha$ will be $u_1^\alpha=(0,u_1^i)$  to
satisfy the constraint, $u^{\prime\alpha}u^\prime_{\alpha}=u^\alpha u_\alpha=-1$
where $u^{\prime\alpha}=u^\alpha+u_1^\alpha$. 
 
To analyze the fate of the perturbation in the dissipative medium 
we assume that the space time dependence of the 
perturbation is  $\sim exp[-i(k  x -\omega t)]$. The perturbations  in different 
components of $T^{\lambda\mu }$ appear as follows\ref{appa}:
\begin{equation}
\begin{aligned}
T^{00}_1&= \epsilon_1 \\  
T^{i0}_1&= (\epsilon +P)u^i_1+\frac{\zeta }{3}\frac{nT^2} {\epsilon +P}\chi \alpha_0 u^i_1\nabla^2\alpha+2\alpha_1\eta \chi \frac{nT^2}{P+\epsilon}\\&\times \{(\vec{u_1}\cdot \vec{\nabla })\del^i\alpha  -\frac{1}{3}u^i_1\nabla ^2\alpha\}\\
    T^{ij}_1&= P_1g^{ij}-\frac{1}{3}\zeta \{i\vec{k}\cdot \vec{u_1}-\frac{1}{3}\zeta\beta_0\omega(\vec{k}\cdot\vec{u_1})\}g^{ij}\\&+\frac{1}{3}\zeta\alpha_0\chi\Big[i\{(\vec{k}\cdot\vec{\nabla })\alpha\}\aleph+\aleph\nabla^2\alpha +\frac{nT^2}{P+\epsilon}\{2(\vec{u_1}\cdot\vec{\nabla})\dot{\alpha}\\&+i\dot{\alpha}(\vec{k}\cdot\vec{u_1}) - i\omega(\vec{u_1}\cdot\vec{\nabla})\alpha\}\Big]g^{ij} -i\eta [k^ju_1^i+k^iu_1^j\\&-\frac{2}{3}g^{ij}(\vec{k}\cdot\vec{u_1})]+2\eta^2\beta_2\omega \{k^ju_1^i +k^iu_1^j-\frac{2}{3}g^{ij}(\vec{k}\cdot\vec{u_1})\}
  \\&+i\alpha _1\eta \chi\aleph[k^j  (\del^i\alpha)   +k^i (\del ^j\alpha) -\frac{2}{3}g^{ij} k_l (\del ^l\alpha) ]\\&+2\alpha_1\eta \chi\frac{nT^2}{P+\epsilon}\{u_1^i(\del^j\dot{\alpha})+u_1^j(\del^i\dot{\alpha})-\frac{2}{3}g^{ij}(\vec{u_1}\cdot\vec{\nabla})\dot{\alpha}\}\\&+i\frac{nT^2}{P+\epsilon}\alpha_1\eta\chi\{k^ju_1^i\dot{\alpha }+k^iu_1^j\dot{\alpha } -\frac{2}{3}g^{ij} (\vec{k}\cdot \vec{u_1})\dot{\alpha}\}
  \end{aligned}
  \label{eq7}
  \end{equation}
 where 
\begin{equation}
\begin{aligned}
\aleph&=\{\frac{n_1T^2+2TT_1n}{P+\epsilon}-\frac{nT^2(P_1+\epsilon_1)}{(P+\epsilon)^2}\}
\label{eq8}
\end{aligned}
\end{equation}
The equations of motions (EoMs) of perturbations dictated by the 
conservations of energy-momentum and net-charge of the fluid 
are given by,
\begin{equation}
\del_\mu T^{\mu\lambda }=0, \,\,\,\,\, \del_\mu N^\mu=0
\label{eq9}
\end{equation} 
The EoMs in the frequency-wave vector space take the 
following form:
\begin{equation}
\begin{aligned}
 0&= \omega T_1^{i0}-k_jT^{ij}_1\\&= \omega (\epsilon +P)u^i_1+\frac{1 }{3}\frac{nT^2} {\epsilon +P}\zeta\omega\chi \alpha_0 u^i_1\nabla^2\alpha   +2\alpha_1\eta \chi\omega \frac{nT^2}{P+\epsilon}\\&\times\{(\vec{u_1}\cdot \vec{\nabla })\del^i\alpha  -\frac{1}{3}u^i_1\nabla ^2\alpha\} -k^iP_1  +\frac{1}{3}\zeta k^i \{i\vec{k}\cdot \vec{u_1} \\&-\frac{1}{3}\zeta\beta_0\omega(\vec{k}\cdot\vec{u_1})\}    -\frac{1}{3}\zeta\alpha_0\chi k^i \Big[i\aleph(\vec{k}\cdot\vec{\nabla })\alpha +\aleph\nabla^2\alpha \\&+\frac{nT^2}{P+\epsilon}  \{2(\vec{u_1}\cdot\vec{\nabla})\dot{\alpha} +i\dot{\alpha}(\vec{k}\cdot\vec{u_1}) -i\omega(\vec{u_1}\cdot\vec{\nabla})\alpha\} \Big]\\& +i\eta [k^2u_1^i +k^i(\vec{k}\cdot\vec{u_1})  -\frac{2}{3}k^i(\vec{k}\cdot\vec{u_1})]-2\eta^2\beta_2\omega \{k^2u_1^i \\&+k^i(\vec{k}\cdot\vec{u_1}) -\frac{2}{3}k^i(\vec{k}\cdot\vec{u_1})\}-i\alpha _1\eta \chi\aleph[k^2(\del^i\alpha) \\& +k^i (\vec{k}\cdot\vec{\nabla})\alpha  -\frac{2}{3} k^i (\vec{k}\cdot\vec{\nabla})\alpha] -\frac{nT^2}{P+\epsilon}2\alpha_1\eta \chi[u_1^i(\vec{k}\cdot\vec{\nabla})\dot{\alpha} \\&+(\vec{k}\cdot\vec{u_1})\del^i\dot\alpha -\frac{2}{3}k^i(\vec{u_1}\cdot\vec{\nabla})\dot{\alpha}]-i\frac{nT^2}{P+\epsilon}\alpha_1\eta\chi\{k^2u_1^i\dot{\alpha }\\&+k^i(\vec{k}\cdot\vec{u_1})\dot{\alpha}    -\frac{2}{3}k^i(\vec{k}\cdot\vec{u_1})\dot{\alpha}\}    
\end{aligned}
\label{eq10}
\end{equation}
The other components of the energy momentum tensor satisfies,
\begin{equation}
\begin{aligned}
0&=  \omega T_1^{00}-k_iT^{i0}_1 \\&  = \omega \epsilon_1- (\epsilon +P)(\vec{k}\cdot \vec{u_1})-\frac{ 1}{3}\zeta\chi \alpha_0\frac{nT^2} {\epsilon +P} (\vec{k}\cdot \vec{u_1}) \nabla^2\alpha  \\& -2\alpha_1\eta \chi \frac{nT^2}{P+\epsilon}\{(\vec{u_1}\cdot \vec{\nabla })(\vec{k}\cdot \vec{\nabla })\alpha  -\frac{1}{3}(\vec{k}\cdot \vec{u_1})\nabla ^2\alpha\} 
\end{aligned}
\label{eq11}
\end{equation}
and  the number conservation equation gives,
\begin{equation}
\begin{aligned}
0&= \omega n_1-n(\vec{k}\cdot \vec{u_1})
\end{aligned}
\label{eq12}
\end{equation}
$P_1$ and $\epsilon_1$ can be expressed 
in terms of  the independent variables, $n_1$ and $T_1$ as follows:
\begin{equation}
\epsilon_1=\Big(\frac{\del\epsilon}{\del T}\Big)_nT_1+\Big(\frac{\del \epsilon}{\del n }\Big)_Tn_1
\label{eq13}
\end{equation} 
and 
\begin{equation}
P_1= \Big(\frac{\del P}{\del T}\Big)_nT_1+\Big(\frac{\del P}{\del n }\Big)_Tn_1
\label{eq14}
\end{equation}
We decompose the fluid velocity into directions perpendicular and parallel to
the direction of wave vector, $\vec{k}$ as:\begin{equation}
\vec{u_1}=\vec{u_1}_\bot +\vec{k}(\vec{k}\cdot\vec{u_1})/k^2
\label{eq15}
\end{equation}
The modes propagating along the direction of $\vec{k}$ are called longitudinal 
and those perpendicular to $\vec{k}$ are called transverse modes. 
Inserting Eq.\ref{eq15} in the EoMs  with the help of Eqs.\ref{eq13} and \ref{eq14} and 
collecting the transverse components, we get the dispersion relation for the transverse mode as:
\begin{equation}
\begin{aligned}
\omega^{\perp} =\frac{-ik^2(\eta-\alpha_1\chi\dot{\alpha})+\frac{n  T^2}{ (P+\epsilon)}2 \alpha _1\eta\chi(\vec{k}\cdot \vec{\nabla
})\dot{\alpha } }{[P+\epsilon-2\eta^2\beta_2k^2+ \frac{n  T^2}{3(P+\epsilon)}   \chi(\alpha_0\zeta-2\alpha_1\eta)\nabla^2\alpha ]}
\end{aligned}
\label{eq16}
\end{equation}
In the acausal limit$( \beta_2=\alpha_1=\alpha_0=0)$, Eq.\ref{eq16} reduces to: 
\begin{equation}
\omega^\perp =\frac{-ik^2\eta }{P+\epsilon  }=i\omega^\bot  _{Im}
\label{eq17}
\end{equation} 
which is the result obtained in acausal hydrodynamics\cite{Weinberg71}. 
We observe that $\zeta$ does not appear in the imaginary part of $\omega^\bot$  and
it is purely imaginary if $\chi=0$.

The derivation of dispersion relation for the longitudinal component is lengthy and
tedious to derive. The details are given in the appendix\ref{appb}. For  $\chi =0$, the imaginary part of 
the longitudinal component of the dispersion relation is:
\begin{equation}
\omega^\parallel_{Im}=\frac{-k^2\{\frac{1}{3}\zeta +\frac{
4}{3}\eta\} } {2\{(P+\epsilon)-\frac{1}{9}k^2\zeta^2\beta_0-\frac{8}{3}k^2\eta^2\beta_2\}}
\label{eq18}
\end{equation}
In the acausal limit, taking $\beta_0=\beta_2=0$   Eq.\ref{eq18} reduces to \begin{equation}
\omega^\parallel_{Im}=\frac{-k^2\{\frac{1}{3}\zeta +\frac{
4}{3}\eta\} } {2 (P+\epsilon) }
\label{eq19}
\end{equation}
which matches with results of \cite{Weinberg71} for $\chi=0$.
The coefficient of $\zeta$ in  Eq.\ref{eq18} differ from that of the one 
given in \cite{Weinberg71} due to the different numerical coefficient of $\zeta$ 
in $T^{\lambda\mu }$ used here.

The real part of the dispersion for the longitudinal modes turns out to be:
\begin{equation}
\begin{aligned}
&\omega^\parallel_{Re} = - \Big[-\{k^2\Big(\frac{\del \epsilon}{\del T}\Big)_n(\frac{1}{3}\zeta 
+\frac{4}{3}\eta)\}^2-4\{(P+\epsilon)\Big(\frac{\del \epsilon}{\del T}\Big)_n\\&
-\frac{1}{9}k^2\zeta^2\beta_0\Big(\frac{\del \epsilon}{\del T}\Big)_n-
\frac{8}{3}k^2\eta^2\beta_2\Big(\frac{\del \epsilon}{\del T}\Big)_n\}\{k^2n\Big(\frac{\del P}{\del T}\Big)_n\\& 
\times \Big(\frac{\del \epsilon}{\del n}\Big)_T -
k^2n\Big(\frac{\del \epsilon}{\del T}\Big)_n\Big(\frac{\del P}{\del n}\Big)_T-
k^2(P+\epsilon)\Big(\frac{\del P}{\del T}\Big)_n)\}\Big]^{1/2}\\& \Big/\Big[{2\{(P+\epsilon)
-\frac{1}{9}k^2\zeta^2\beta_0
 -\frac{8}{3}k^2\eta^2\beta_2  \}\Big(\frac{\del \epsilon}{\del T}\Big)_n}\Big]
\end{aligned}
\label{eq20}
\end{equation}
In the acausal limit, considering vanishing net number density$(n)$, Eq.\ref{eq20} reduces to:
 
\begin{equation}
\begin{aligned}&\omega^\parallel_{Re} = - \Big[-\{k^2\Big(\frac{\del \epsilon}{\del T}\Big) (\frac{1}{3}\zeta +\frac{4}{3}\eta)\}^2   -4(P+\epsilon)\Big(\frac{\del \epsilon}{\del T}\Big)    \\&\times  \{-k^2(P+\epsilon)\Big(\frac{\del P}{\del T}\Big) )\}\Big]^{1/2} \Big/\Big[ {2\{(P+\epsilon)\Big(\frac{\del \epsilon}{\del T}\Big)    \}}\Big]
\end{aligned}
\label{eq21}
\end{equation}
which appears as:
\begin{equation}
\omega_{Re}=c_s|k|+(constant)k^2
\label{eq22}
\end{equation}
where $c_s$ is the speed of sound wave in the fluid. 
The acausal limit $\omega_{Re}=c_s|k|$ can be recovered 
by keeping only the linear term~\cite{Weinberg71}. The causal dispersion relation derived
here can reproduce all the known relations exist in the acausal limit.

\section{Effects of magnetic field}
\label{sec:3}
It has been shown that a ultra-high but transient magnetic field 
is generated in the collision of heavy ions at RHIC and LHC energies
~\cite{skokov}. Survivability of the magnetic field will depend on the value
of the conductivity of the QGP medium formed in these collisions.  The presence of magnetic field
will affect the properties of the fluid through its contribution to the energy-momentum tensor. 
Considering constant magnetic field($B$), the magnetic contribution is given 
by\cite{gedalin}
\begin{equation}
T^{\mu\nu}_{m}=\frac{B^2}{8\pi }(2u^\mu u^\nu+g^{\mu\nu}-2n'^\mu n'^\nu )
\label{eq23}
\end{equation} where $n'^\mu $ is the unit vector in the direction of the magnetic field $n'^\mu=B^\mu/B $ with $n'^\mu n'_\mu=-1 $ and $ u^\mu n'_\mu=0$
The conservation equation then reads: 
\begin{equation}
\del _\mu T^{\mu \nu}_{tot}=\del _\mu T^{\mu \nu}+\del _\mu T^{\mu \nu}_{m}=0
\label{eq24}
\end{equation}
where 
\begin{equation*}
T^{\mu \nu}_{tot}=  T^{\mu \nu}+  T^{\mu \nu}_{m}
\end{equation*}
For small perturbation, $T^{\mu\nu}_{1m}$ to $T^{\mu\nu}_m$ 
the first  and third term of Eq.\ref{eq23} will be changed.
The changes in different components of $T^{\mu \nu}_{1,m}$ are given in Appendix\ref{appc}.
 After taking Fourier transformation, the equations of motions  in the presence of magnetic field becomes
 \begin{equation}
 \begin{aligned}
  0&=\omega T_{1,tot}^{00}-k_iT^{i0}_{1,tot}\\& =F_1-\frac{B^2}{8\pi }[2(\vec{k}\cdot\vec{u_1})-\frac{(\vec{B}\cdot\vec{k})}{B^2}(\vec{B}\cdot\vec{u_1})]
 \label{eq25}
 \end{aligned}
 \end{equation} 
 \begin{equation}
  \begin{aligned}
    0&= \omega T_{1,tot}^{i0}-k_jT^{ij}_{1,tot}\\&=F_2+\frac{\omega B^2}{ 8\pi } [2 u_1^i-\frac{B^i}{B^2}(\vec{B}\cdot\vec{u_1}]
  \label{eq26}
  \end{aligned}
  \end{equation} 
 where $F_1$ and $F_2$ are the expressions given in the right hand side of Eqs.~\ref{eq10} and ~\ref{eq11} respectively.
 Decomposing the fluid velocity as Eq.\ref{eq15}, Eq\ref{eq25} becomes
 \begin{equation}
 0 =F^\prime_1-\frac{B^2}{8\pi }\Big[2(\vec{k}\cdot\vec{u_1})-\frac{(\vec{B}\cdot\vec{k})}{B^2
 }(\vec{B}\cdot\vec{u}_{1\perp})-\frac{(\vec{B}\cdot\vec{k})^2}{B^2k^2}(\vec{k}\cdot\vec{u_1})\Big]
  \label{eq27} 
 \end{equation}
 If we take constant magnetic field 
along the direction of wave vector $k$, then $(\vec{B}\cdot\vec{k})=kB$ and $(\vec{B}\cdot\vec{u}_{1\perp})=0$ as $\vec{u}_{1}\perp \vec{k}.$  In that case Eq.\ref{eq27} becomes,
 \begin{equation}
  0 =F^\prime_1-\frac{B^2}{4\pi }  (\vec{k}\cdot\vec{u_1}) 
   \label{eq28} 
   \end{equation}
   Similarly, after decomposition Eq.\ref{eq26} reads as:
  \begin{equation}
   \begin{aligned}
     0&=  F^\prime_2+\frac{\omega B^2}{ 8\pi } [2 u_{1\perp}^i+\frac{2k^i}{k^2}(\vec{k}\cdot\vec{u_1})-\frac{B^i}{Bk}(\vec{k}\cdot\vec{u_1)}]
   \label{eq29}
   \end{aligned}
   \end{equation}
   and the number conservation equation remains unchanged. Here $F^\prime_1$ and $F^\prime_2$ represents 
the right hand side (RHS) of Eq.\ref{eq10} and \ref{eq11} respectively after decomposition of fluid velocity. 
The dispersion relation in the transverse direction in the presence of constant $B$ is given by,
      \begin{equation}
      \begin{aligned}
      \omega^{\perp} =\frac{-ik^2(\eta-\alpha_1\chi\dot{\alpha})+\frac{n  T^2}{ (P+\epsilon)}2 \alpha _1\eta\chi(\vec{k}\cdot \vec{\nabla
      })\dot{\alpha } }{[P+\epsilon-\frac{B^2}{4\pi}-2\eta^2\beta_2k^2+ \frac{n  T^2}{3(P+\epsilon)}   \chi(\alpha_0\zeta-2\alpha_1\eta)\nabla^2\alpha ]}
      \end{aligned}
      \label{eq30}
      \end{equation}
      For $B=0$, Eq.\ref{eq30} reduces to Eq.\ref{eq16}.
       The imaginary part of the  dispersion relation in  the longitudinal direction 
can be expressed as, 
      \begin{equation}
            \begin{aligned}
            \omega^{\parallel}_{Im} =\frac{-k^2(\frac{1}{3}\zeta+\frac{4}{3}\eta)  }{2[P+\epsilon+\frac{B^2}{8\pi} -\frac{1  }{9}k^2\zeta^2\beta_0-\frac{8}{3}k^2\eta^2\beta^2]}
            \end{aligned}
            \label{eq31}
            \end{equation}
and the real part is:
      \begin{equation}
      \begin{aligned} \omega^\parallel_{Re}&= \Big[-\Big\{k^2\Big(\frac{\del \epsilon}{\del T}\Big)_n(\frac{1}{3}\zeta +\frac{4}{3}\eta)\Big\}^2   -4\Big\{(P+\epsilon)+\frac{B^2}{8\pi}\\& -\frac{1}{9}k^2\zeta^2\beta_0    -\frac{8}{3}k^2\eta^2\beta_2\Big\}\Big(\frac{\del \epsilon}{\del T}\Big)_n \Big\{k^2n\Big(\frac{\del P}{\del T}\Big)_n\Big(\frac{\del \epsilon}{\del n}\Big)_T \\& -\frac{k^2 B^2}{4 \pi }\Big(\frac{\del P}{\del T}\Big)_n-k^2n\Big(\frac{\del \epsilon}{\del T}\Big)_n\Big(\frac{\del P}{\del n}\Big)_T  \\&-k^2(P+\epsilon)\Big(\frac{\del P}{\del T}\Big)_n\Big\}\Big]^{1/2}\\& \Big/\Big[ 2\Big\{(P+\epsilon)+\frac{B^2}{8\pi} -\frac{1}{9}k^2\zeta^2\beta_0   -\frac{8}{3}k^2\eta^2\beta_2 \Big  \}\Big(\frac{\del \epsilon}{\del T}\Big)_n \Big]
      \end{aligned}
      \label{eq32}
      \end{equation}
      where we have considered $\chi =0$  to keep the expression compact, however, the derivation of the dispersion relation 
      for $\chi\neq 0$ is straight forward.
\section{Effects of Causality on Fluidity}
\label{sec:4}
What is the difference that it makes to characterize a relativistic fluid by
using  causal vis-a-vis acausal dispersion relations? 
In the following we will study this aspect in details. The fluidity of QGP can be 
studied ~\cite{koch_prc} by introducing  the ratio of two length scales -
one of those is related to the wave length of the sound wave propagating 
through the fluid. The other one is the inter-particle distance in the fluid. 

\subsection{Viscous horizon}
In the following we will provide the threshold value of wave vector, $k_v$ above which no sound wave 
can propagate. The quantity, $R_v\sim k_v^{-1}$ determines the 
length scale called viscous horizon~\cite{staig}.
The imaginary part of dispersion relation dictates the attenuation of sound wave in the fluid.
A sound wave damps in time as $\sim \exp(\omega_{Im}t)$  (for $\omega_{Im}<0$) in viscous medium.
This can be expressed in terms of perturbation to $T^{\mu\nu }$ as:
\begin{equation}
T_1^{\mu\nu }(t)=T_1^{\mu\nu }(t_i)\exp(   \omega_{Im} t)
\label{eq33}
\end{equation}
where $T_1^{\mu\nu }(t_i)$ represents the perturbation to $T^{\mu\nu }$ at the initial time $t_i$.
The dispersion relation derived in the previous section may be used to determine
upper limit of wave vector $k_v$ of the sound wave that can propagate in the medium,
which can be obtained by setting:  $|\omega_{Im}|t=1$
\begin{equation}  
k^{ causal}_{v, long}\equiv\frac{1}{R_{v,long}^{causal}}=
\sqrt{\frac{P+\epsilon}{\frac{t}{2}(\frac{\zeta}{3}+\frac{4\eta}{3})+\frac{1}{9}\zeta^2\beta_0+\frac{8}{3}\eta^2\beta _2}}
\label{eq34}
\end{equation}
We note that the viscous horizon scale, $R_v\sim \sqrt{t}$ in contrast
to sound horizon which varies linearly with $t$.
The above condition implies that a longitudinal mode  with magnitude of $k$ 
larger than $k^{causal}_{v,long}\equiv 1/{R_{v,long}^{causal}}$ will be killed by
dissipation and all other longitudinal modes with lower values of $k$ will propagate. 
The known result~\cite{staig} in the acausal limit ($\beta_0=\beta_2=0$) can be obtained as:
\begin{equation}  
k^{acausal}_{v,long}
\equiv\frac{1}{R_{v,long}^{acausal}}=
\sqrt{\frac{P+\epsilon}{\frac{t}{2}(\frac{\zeta}{3}+\frac{4\eta}{3})}}
\label{eq35}
\end{equation} 
Similarly for the causal transverse mode   we have the upper limit,
\begin{equation}  
k_{v,tran}^{causal}\equiv\frac{1}{R_{v,tran}^{causal}}=\sqrt{\frac{P+\epsilon}{\eta(t+2\eta\beta_2)}}
\label{eq36}
\end{equation} 
and  in the acausal limit the above relation turns out to be 
\begin{equation}  
k_{v,tran}^{acausal}
\equiv\frac{1}{R_{v,tran}^{acausal}}=
\sqrt{\frac{P+\epsilon}{\eta t}}
\label{eq37}
\end{equation}
We have already seen in the previous section that the 
application of magnetic field changes the  
dispersion relations. Therefore, the viscous horizon in presence
of magnetic field should also change to: 
\begin{equation}  
k_{v,tran,B}^{causal}\equiv\frac{1}{R_{v,tran,B}^{causal}}=\sqrt{\frac{P+\epsilon-\frac{B^2}{4\pi }}{\eta(t+2\eta\beta_2)}}
\label{eq38}
\end{equation} 
\begin{equation}  
k_{v,long,B}^{causal}\equiv\frac{1}{R_{v,long,B}^{causal}}=
\sqrt{\frac{P+\epsilon+\frac{B^2}{8\pi}}{\frac{t}{2}(\frac{\zeta}{3}+\frac{4\eta}{3})+\frac{1}{9}\zeta^2\beta_0+\frac{8}{3}\eta^2\beta _2}}
\label{eq39}
\end{equation} 
The viscous horizon has an impact on the flow harmonics. It is argued in~\cite{lacey} that the
properties related to the ratio of higher order to second order harmonics, {\it i.e.} $v_n/v_2$ with ($n>2$) 
can be understood in terms of the propagation of sound wave through dissipative medium and hence such studies
will help in estimating the size of the sound horizon and viscous horizon~\cite{staig}. 
 
\subsection{Measure of fluidity}
Sound wave in a viscous fluid will stop propagating if its wave length is smaller
than some theroshold value, $\lambda_{th}=2\pi/k_v$.  The value of $\lambda_{th}$ will depend 
on the values of dissipative coefficients, $\eta$, $\zeta$,  $\chi$, etc.
The fluidity of the system has been defined in Ref.~\cite{koch_prc,VK}  with
the introduction of a new quantity which depends on the intrinsic properties 
of the fluid and enables one to compare fluids of wide varieties 
such as non-relativistic fluid like water and relativistic, extremely dense and hot fluid like QGP.
For example, the temperature of water and QGP differ by a factor $\sim O(10^{10})$.
Now if we want to compare their fluidity we may find the dissipation per inter-particle
separation. 
In Ref.~\cite{koch_prc} the linearized first order dispersion relation of the sound mode was used,
\begin{equation} 
\omega=c_s k -\frac{i}{2}k^2 \frac{\frac{4}{3}\eta}{h/c^2}
\label{eq40}
\end{equation} 
The imaginary part of the dispersion relation represents the dissipation of 
sound wave in the medium. The sound mode with wave vector $k$ will propagate if the imaginary part of frequency 
is small {\it i.e.}:
\begin{equation}
\Big|\frac{\omega_{Im}(k)}{\omega_{Re}(k)}\Big|\ll 1
\label{eq41}
\end{equation}
The limiting value can be found by setting $\mid{\omega_{Im}}/{\omega_{Re}} \mid =1$, 
which gives $k=3hc_s/(2\eta)$ then the resulting threshold for wavelength of the sound mode becomes
\begin{equation}
\lambda_{th}=\frac{2\pi}{k_v}=\frac{4\pi}{3}\frac{\eta}{h c_s}=\frac{4\pi}{3} L_\eta
\label{eq42}
\end{equation}
where $L_\eta=\eta/(hc_s)$. The $L_\eta$ gives an estimation for lowest sound 
wavelength ($\lambda_{th}$) which can propagate through the viscous fluid.
The $L_\eta$ has the dimension of length and can be used to characterize fluids. 
However, introduction of a dimensionless scale 
will enable us to compare fluids with varying densities. 
Quantities like Reynolds  or Knudsen numbers have been used in Refs.~\cite{Bonasera} and
~\cite{Gombeaud} respectively to study flow properties. However, both of these quantities
involve parameters, like dimension of the system which is not connected with the intrinsic
properties of the fluid. 
The particle number density($\rho$) can be used to estimate the inter-particle distance,
$L_{\rho}\sim \rho^{-1/3}$, which is related to the intrinsic properties of the fluid.
The ratio of $L_\eta$ to $L_\rho$ may be used to characterize the fluid.
For relativistic QGP with vanishing net baryon number density, entropy density ($s$)
can be used to estimate $\rho$ by using $\rho\sim s/4$.
The ratio of these two length scales can be used as a measure of fluidity
\begin{equation}
F\equiv\frac{L_\eta}{L_\rho}
\label{eq43}
\end{equation}

What is corresponding expression of $F$  for causal fluid dynamics involving other
transport coefficients in addition to $\eta$?  
We use dispersion relations derived from causal relativistic 
hydrodynamics involving shear, bulk viscosities, thermal conductivity and different relaxation coefficients
to estimate the fluidity. We would contrast our results to those obtained with
acausal relation~\cite{koch_prc}. The length scale analogous to $L_\eta$ for causal fluid 
dynamics is denoted by $L_T$
depends on the transport coefficients like $\zeta$, $\chi$, $\beta_0$,$\beta_2$ in addition to $\eta$.
$L_T$ for the longitudinal mode is given by,
\begin{equation}
\begin{aligned} 
L_T &=\Big [\frac{1}{4}\Big]\Big\{\Big(\frac{\del \epsilon}{\del T}\Big)_n \zeta^2+8\zeta\eta\Big (\frac{\del \epsilon}{\del T}\Big)_n+16\eta^2\Big(\frac{\del \epsilon}{\del T}\Big)_n\\&+2(P+\epsilon)\beta_0\zeta^2\Big(\frac{\del P}{\del T}\Big)_n+2n\beta_0\zeta^2\Big(\frac{\del \epsilon}{\del T}\Big)_n\Big(\frac{\del P}{\del n}\Big)_T\\&-2n\zeta^2\beta_0\Big(\frac{\del P}{\del T}\Big)_n\Big(\frac{\del \epsilon}{\del n}\Big)_T   +48 \eta^2\beta_2(P+\epsilon)\Big(\frac{\del P}{\del T}\Big)_n\\& +48\eta^2\beta_2n\Big(\frac{\del \epsilon}{\del T}\Big)_n\Big(\frac{\del P}{\del n}\Big)_T-48\eta^2\beta_2n\Big (\frac{\del P}{\del T}\Big )_n\Big(\frac{\del \epsilon}{\del n}\Big)_T\Big\}^\frac{1}{2}\\&\Big/\Big\{(P+\epsilon)^2\Big(\frac{\del P}{\del T}\Big)_n+n(P+\epsilon)\Big(\frac{\del \epsilon}{\del T}\Big)_n\Big(\frac{\del P}{\del n}\Big)_T\\&-n(P+\epsilon)\Big(\frac{\del \epsilon}{\del n}\Big)_T\Big(\frac{\del P}{\del T}\Big)_n\Big\}^{1/2}
\label{eq44}
\end{aligned}
\end{equation}
We use $({\del P}/{\del T})=({\del P}/{\del\epsilon})({\del \epsilon}/{\del T})$ to express $F$ as:
\begin{equation}
\begin{aligned} 
F&=\Big[\frac{{\rho }^\frac{1}{3}}{4}\Big]\Big \{  \zeta^2+8\zeta\eta  +16\eta^2 +2(P+\epsilon)\beta_0\zeta^2 \Big(\frac{\del P}{\del\epsilon} \Big)_n\\&+2n\beta_0\zeta^2 \Big(\frac{\del P}{\del n}\Big)_T-2n\zeta^2\beta_0 \Big(\frac{\del P}{\del \epsilon}\Big)_n \Big(\frac{\del \epsilon}{\del n}\Big)_T   +48 \eta^2\beta_2\\&\times (P+\epsilon)  \Big(\frac{\del P}{\del \epsilon}\Big )_n +48\eta^2\beta_2 n\Big(\frac{\del P}{\del n}\Big)_T-48\eta^2\beta_2n\Big (\frac{\del P}{\del \epsilon}\Big )_n\\&\times  \Big (\frac{\del \epsilon}{\del n}\Big )_T\Big\}^\frac{1}{2} \Big /\big \{(P+\epsilon)^2\Big (\frac{\del P}{\del \epsilon}\Big)_n+n(P+\epsilon)\Big (\frac{\del P}{\del n}\Big )_T\\&-n(P+\epsilon)\Big(\frac{\del \epsilon}{\del n}\Big)_T\Big(\frac{\del P}{\del T}\Big)_n\big \}^{1/2}
\end{aligned}
\label{eq45}
\end{equation}
This is measure of fluidity of a relativistic fluid for $\chi =0$.

For a fluid having vanishing net charge density ($n=0$) the above equation becomes
\begin{equation}
\begin{aligned}
& F=\Big[\frac{{\rho }^\frac{1}{3}}{4} \Big \{\zeta^2+8\zeta\eta+
16\eta^2 +2(P+\epsilon)\beta_0\zeta^2\Big(\frac{\del P}{\del \epsilon}\Big )\\& 
+48 \eta^2\beta_2(P+\epsilon)\Big(\frac{\del P}{\del\epsilon}\Big)\Big\}^\frac{1}{2}\Big]\Big/\Big\{(P+\epsilon)^2\Big(\frac{\del P}{\del\epsilon}\Big)\Big\}^{1/2}
\label{eq46}
\end{aligned}
\end{equation}
It is clear from this result that  dispersion relations become more 
complex if relativistic causal hydrodynamics is used. Two more coefficients,
$\beta_0$ and $\beta_2$ enter into the expression for fluidity. In the acausal limit i.e. for vanishing 
$\beta_0$ and $\beta_2$ as well as neglecting non linear terms in the real part of $\omega$, 
the $F$ reads,
\begin{equation}
F=\frac{\rho^{1/3}\eta}{h c_s}
\label{eq47}
\end{equation}  
which is exactly what is given in Ref.\cite{koch_prc}.  
It may be noted from Eq.~\ref{eq45} that the fluidity measure, $F$ of the causal fluid has 
a complicated functional dependence on 
various transport coefficients and thermodynamic variables of the fluid.
In contrast to the  causal case the  $F$  has simpler dependence on
transport coefficients and thermodynamical variables in an acausal scenario (Eq.~\ref{eq47}).

\subsection{Fluidity in presence of magnetic field} 
We have already seen that non-zero $B$ affects the real and imaginary part of $\omega$ along the longitudinal direction and hence 
it modifies the fluidity measure also. For vanishing net charge and $\zeta=\chi=0$ the $L_T$ becomes,
\begin{equation}
\begin{aligned}   
L_T&=\Big[ \Big(\frac{\del\epsilon}{\del P}\Big)(\zeta+4\eta)^2 +2\beta_0\zeta^2(\frac{B^2}{4\pi }+P+\epsilon)\\&+48\eta^2\beta_2(\frac{B^2}{4\pi }+P+\epsilon) \Big]^{1/2}\Big/\Big[4\Big\{2\Big(\frac{B^2}{8\pi }\Big)^2\\&+ \frac{3B^2}{8\pi }(P+\epsilon)+(P+\epsilon)^2\Big \}^{1/2}\Big]
\label{eq48} 
\end{aligned}
\end{equation}
For simplicity we kept only $\eta$ as non-zero. However,
it is straight forward to find $F$ with  non-zero $n,\zeta$ and $\chi$.
\section{Results and Discussion}
\label{sec:5}
In this section we discuss the dispersion relation for the transverse and
longitudinal modes for non-expanding fluid. 
\subsection{Transverse mode}
In order to see how causality or causal hydrodynamics 
affects the damping of sound wave, first we consider 
the transverse component of the dispersion relation. For $\chi =0$, Eq.~\ref{eq16} reads:
\begin{equation}
\begin{aligned}
\omega^\bot _{Im} =\frac{-k^2 \eta }{[P+\epsilon-2\eta^2\beta_2k^2 ]}
\label{eq49}
\end{aligned}
\end{equation}
It is interesting to note that the bulk viscosity does not appear in the 
dispersion relation for the transverse mode. 
The coefficient $\beta_2$ appearing in the denominator is the signature of causal hydrodynamics. 
In the ultra-relativistic limit it has the limiting value\cite{Israel79}
\begin{equation}
\beta_2=\frac{3}{4P}
\label{eq50}
\end{equation} 
We  estimate the damping of the  sound wave by 
using the thermodynamic relation for vanishing net charge density 
(such as baryon free  QGP),  $P+\epsilon=sT$. 
\begin{figure}
\begin{center}
\includegraphics[scale=0.95]{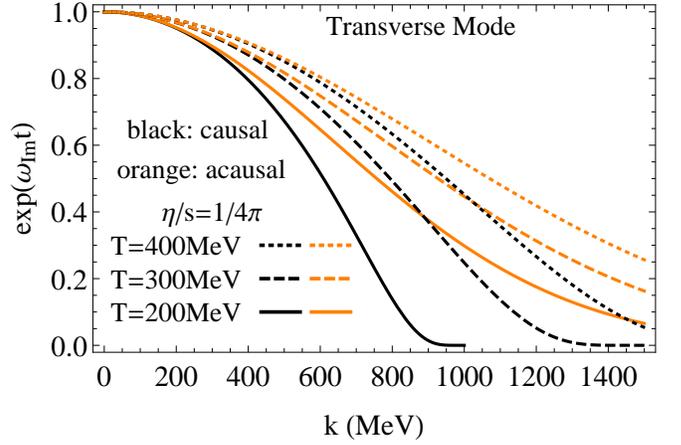}
\caption{(colour online) Damping  of transverse perturbative modes in QGP  with $k$   
at $T=200, 300$ and $400$ MeV for causal and acausal hydrodynamics. We have taken $t=0.6$ fm/c}
\label{f1}
\end{center}
\end{figure}
In Fig.~\ref{f1} we display the damping of the transverse mode
with $k$ for $\eta/s=1/4\pi$ at $T=$200, 300 and 400 MeV. 
We find that damping is stronger for 
larger $\eta/s$, lower $T$ and larger wave numbers or smaller  wave lengths. 
The imaginary part of the dispersion
relation leads to the variation of amplitude with $k$ as $\sim \exp(-\Gamma_s k^2)$ where
the $\Gamma_s$, square of the characteristic dissipation length  that
picks up different values at causal and acausal scenario resulting
in different damping rate for different $k$.
Although for small $k$ it is not significant but at large $k>200$ MeV the  difference 
is distinctly visible in the results displayed in Fig.~\ref{f1}. 
\begin{figure}
\begin{center}
\includegraphics[scale=0.92]{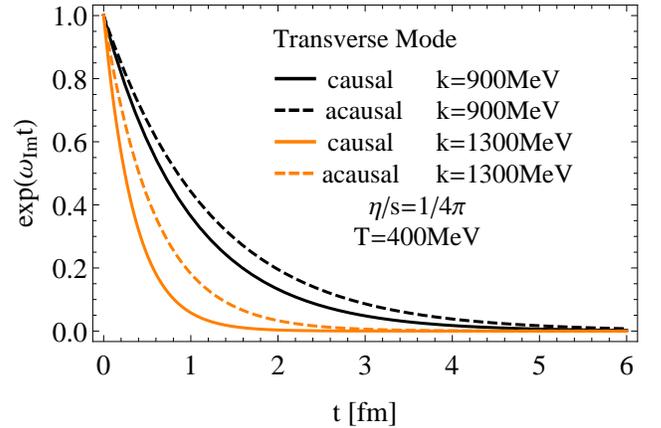}
\caption{(colour online) Damping of the transverse perturbative modes in QGP with time (t)  at $T=400$ MeV for 
$\eta/s=1/4\pi$ for causal and acausal hydrodynamics for various $k$.} 
\label{f2}
\end{center}
\end{figure} 
The decay of the perturbation with time is shown in Fig.~\ref{f2} for $\eta /s=1/4\pi$ 
for different $k$. 
We observe that at $T=400$ MeV the perturbations decay faster in causal than acausal hydrodynamic  
as $k$ increases. Stronger damping  is observed at $T=200$ and 300 MeV (not shown in the figure). 
At large $t$, the amplitude of the perturbations for causal and acausal scenarios are close  because
at large $t$ the amplitude decays to a very small value irrespective of the value of $\omega_{Im}$. Similarly at
small $t$ the amplitude of the perturbation are also close.
The enhanced magnitude of $\eta/s$ enforces  faster  decay. 
All these results represent a physically consistent picture because it is well-known that 
in the acausal (first order) hydrodynamics a non-equilibrium  system evolves to the 
equilibrium instantly. 
However,  in second order hydrodynamics the non-equilibrium system does not go to the 
equilibrium state instantaneously but takes some non-zero time. This non-zero time lag is incorporated 
in the second order hydrodynamics by means of relaxation coefficients such as $\beta_0, \beta_1, 
\beta_2$. In other words the second order hydrodynamics effectively enhances the 
dissipation of the system. As any disturbance will dissipate faster in a higher order viscous hydrodynamics than the 
lower one, the perturbations in causal disturbances fall faster than the acausal one.  
We have observed that the amplitude of sound wave falls faster with increase in $\eta/s$ and decrease in $T$.

\subsection{Longitudinal mode}
\begin{figure}
\begin{center}
\includegraphics[scale=0.95]{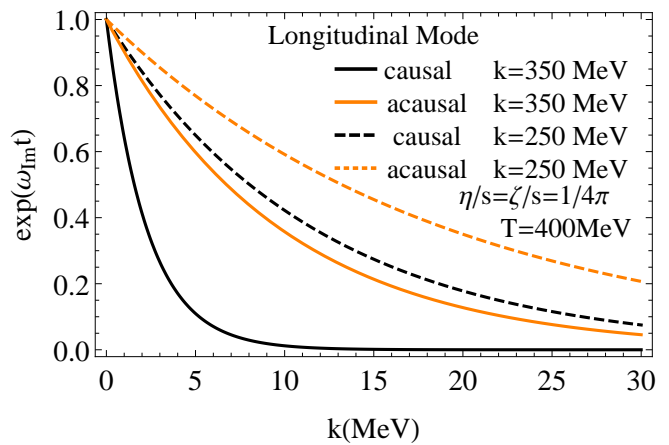}
\caption{(colour online) Damping of the longitudinal mode  with time (t)  at $T=400$ 
MeV for $\eta/s=1/4\pi$ and $\zeta/s=1/4\pi$ for various $k$ values.} 
\label{f4}
\end{center}
\end{figure}

\begin{figure}
\begin{center}
\includegraphics[scale=0.85]{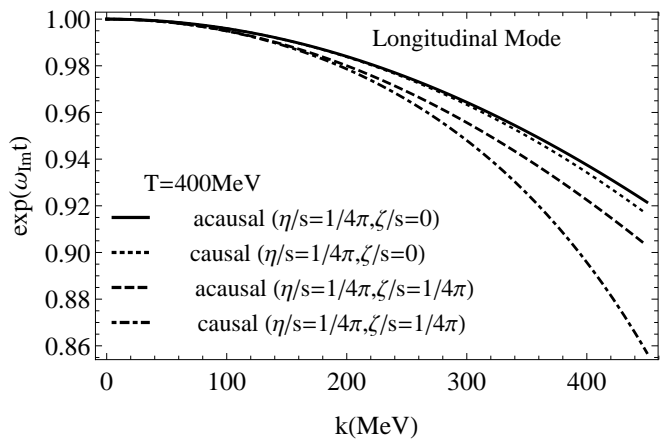}
\caption{Damping of the perturbations in QGP for different values of $\eta/s$ and $\zeta/s$.
$t$ is taken as 0.6 fm/c } 
\label{f5}
\end{center}
\end{figure}

To study the perturbations in longitudinal direction, we encounter a new relaxation coefficient, $\beta_0$
that was absent in acausal theory.  In the ultra-relativistic limit $\beta_0$ is given by \cite{Israel79},
\begin{equation}
\beta_0=\frac{216}{P\beta^4}
\label{eq51}
\end{equation}
where $\beta = m/T$. We have used thermal mass to estimate $\beta$.
To study the propagation of the longitudinal modes in the fluid
we consider the gluonic fluid.  The thermal mass of gluon is 
given by~\cite{Bellac}
\begin{equation}
\frac{m_g}{T}=g\sqrt{\frac{C_A+N_f/2}{6}}\,\,\,\,\, \Rightarrow \beta=g\sqrt{\frac{C_A+N_f/2}{6}} 
\label{eq52}
\end{equation}
where $g=\sqrt{4\pi\alpha_s}$, $C_A=3$ and $N_f=2$ (for two flavours).
In the present work we have taken $\alpha_s=0.2$.
\begin{figure}
\begin{center}
\includegraphics[scale=0.95]{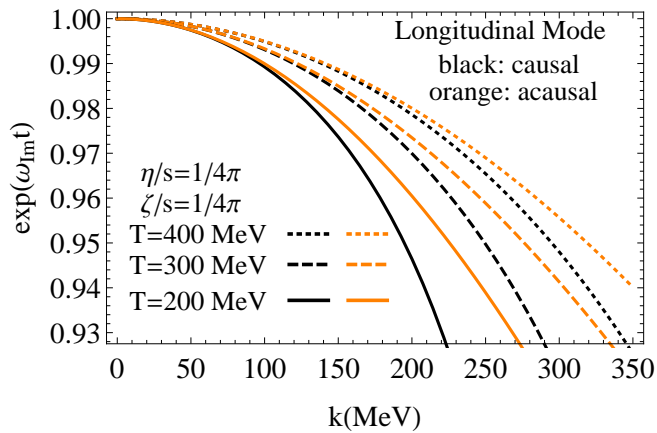}
\caption{(colour online) Damping  of perturbations 
with $k$ in the longitudinal direction for $T=200, 300$ and 400 MeV. 
We have taken $t=0.6$ fm/c.}
\label{f3}
\end{center}
\end{figure} 
We use Eq.~\ref{eq18} with the aid of $\beta_0$ to  study the dissipation of the
longitudinal modes.
One major difference with the transverse mode is the appearance of bulk viscosity in
the longitudinal mode and it will be seen later that bulk viscosity plays dominant 
role in the damping of the perturbations.
The nature of variation of the perturbations of longitudinal mode is similar to that of 
transverse modes. The perturbation decays faster with $k$ in causal than acausal hydrodynamics (Fig.~\ref{f3}).
At lower $T$  a faster decay is observed.
In Fig.~\ref{f4}, we depict the dissipation of the perturbations with time for 
$\eta/s=1/4\pi$ for different $k$ values.
A faster decay is observed at higher $\eta/s$  and lower $T$.
Similar to the transverse modes the difference in the decay of longitudinal
amplitudes in causal and acausal hydrodynamics is significant. 
We have discussed before that the longitudinal dispersion relation is
controlled not only by shear but by the bulk viscosity as well.
The damping  of the longitudinal modes due to shear and bulk viscous coefficients and 
the relative importance  of these coefficients are investigated.
The variation of the damping with $k$ has been depicted in 
Figs.~\ref{f5}. The result indicates a bigger influence of the bulk viscosity
on the longitudinal modes than the shear viscosity.

\begin{figure}
\begin{center}
\includegraphics[scale=0.93]{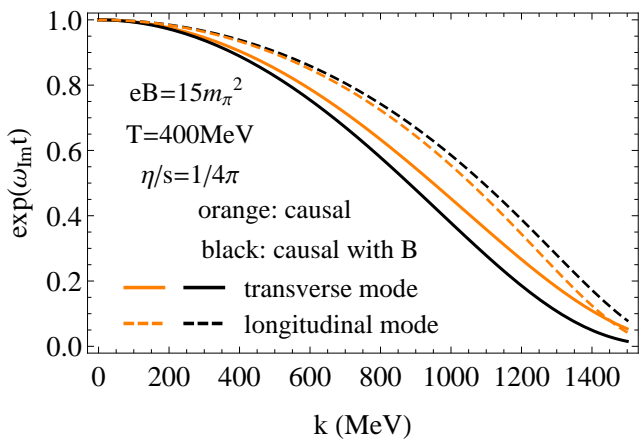}
\caption{(colour online) Damping of perturbations with $k$  in QGP 
in the presence of B along the transverse direction (solid line) and longitudinal direction (dashed line) at T=400 MeV.
The value of $t$ is taken as 0.6 fm/c here.} 
\label{f6}
\end{center}
\end{figure}

\begin{figure}
\begin{center}
\includegraphics[scale=0.95]{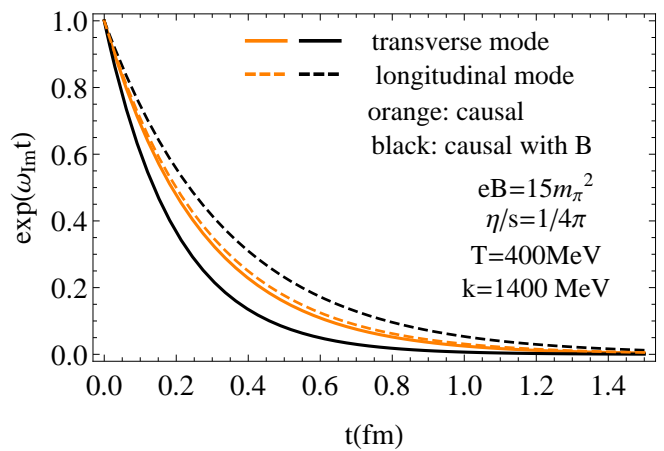}
\caption{(colour online) Time variation of damping  of the transverse mode (solid line) and longitudinal mode (dashed line) at T= 400 MeV in the presence B.} 
\label{f7}
\end{center}
\end{figure}

As mentioned in section \ref{sec:3} the QGP fluid may be subjected to the external magnetic field ($B$) created
due to the relativistic motion of the colliding nuclei.  The magnitude of the field during evolution of  QGP 
will depend on the rate of decay of the field which is controlled
by the value of electrical conductivity of QGP. We assume a non-zero constant  magnetic field 
in the QGP and study its effects on the fluid properties. 
We find that the energy due to magnetic field appears with opposite sign in the denominators of 
$\omega^\perp$ and $\omega^\parallel$ given by Eqs.~\ref{eq30} and ~\ref{eq31} respectively.
This is reflected in the results displayed Figs.~\ref{f6} and ~\ref{f7} for the variation of damping with $k$ and $t$
respectively. The transverse modes decays faster in causal hydrodynamics. An opposite trend is observed
for the longitudinal modes. 

\subsection{Quantitative changes in the viscous horizon}
We would like to estimate the shift in the viscous horizon caused by causal hydrodynamics
as compared to the acausal one. The viscous horizon size scales with time as: $R_v\sim 1/\sqrt{t}$. 
Through the relation, $R_v$(fm)$\approx 197/k_v$(MeV), it determines the wave length
that is unable to  propagate in the dissipative medium, {\it i.e.} if
the wave length is less than $2\pi/k_v$ then those waves will dissipate.

\begin{figure}
\begin{center}
\includegraphics[scale=0.95]{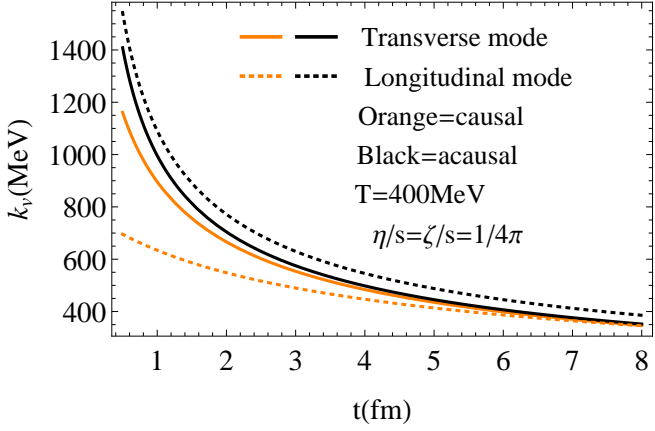}
\caption{(colour online) Variation of  $k_v$ with $t$ for 
causal and acausal hydrodynamics at $T=400$ MeV.
} 
\label{f7a}
\end{center}
\end{figure}

Using Eqs.~\ref{eq34},\,\ref{eq35},~\ref{eq36}\, and\, \ref{eq37} we can estimate 
viscous horizon scales at different time. The variation of $k_v$ with $t$ for 
causal and acausal hydrodynamics has been depicted in Fig.~\ref{f7a} at $T=400$ MeV.  
It is observed that  $k_v$ for the causal scenario approaches 
the $k_v$ for the acausal scenario at large $t$.  This trend can be understood from 
the mathematical expressions of  Eqs.~\ref{eq34} and \ref{eq35}.
However, if the time variation of pressure due to hydrodynamic 
evolution is considered then $\beta_0$ and $\beta_2$ will also increase with time 
as evident from Eqs.~\ref{eq50} and \ref{eq51} and in such situation the difference between the
causal and acausal scenario may survive at large $t$ also. 

\begin{figure}
\begin{center}
\includegraphics[scale=0.95]{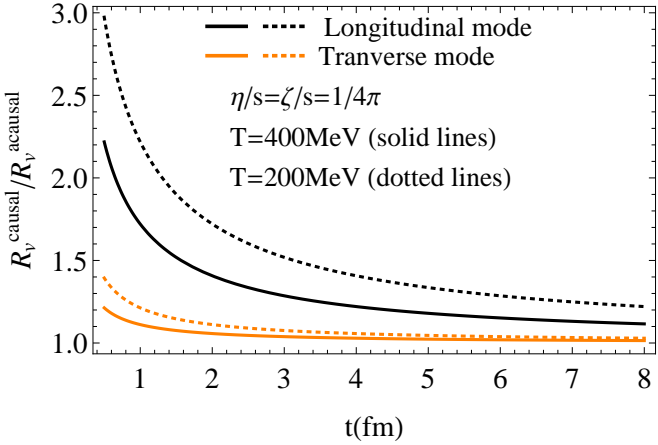}
\caption{(colour online) Variation of the ratio of the viscous horizon 
length ($R_v$) with $t$ for causal and acausal hydrodynamics.
} 
\label{f7b}
\end{center}
\end{figure}

In Fig.~\ref{f7b}, we display the ratio of viscous horizon lengths for causal and acausal
hydrodynamics as a function of $t$ for $T=200$ and 400 MeV. We find that the longitudinal
scale in causal hydrodynamics is almost 3 times larger than acausal one at $t=0.6$ fm for
$T=200$ MeV and $\eta/s=\zeta/s=1/4\pi$. The same ratio becomes 2.07 for $T=400$ MeV at $t=0.6$ fm/c. 
We also note that the difference in the viscous horizon length for transverse  modes
is smaller than the longitudinal modes.

The viscous damping controls the highest order of flow harmonic ($n_v$)
that will survive against the dissipative effects.  The relation between
$n_v$ and $R_v$ is given by ~\cite{lacey}: $n_v=2\pi R/R_v$ where
$R$ is the size of the fluid system. Therefore, an increase in $R_v$ will
reduce the value of $n_v$ resulting in a shift in its value between casual 
and acausal scenarios.  Since the value of $n_v$ depends on $\eta/s$,
measurement of amplitudes of various harmonics will help in 
determining the viscosity and 
consequently characterizating  QGP~\cite{staig}.
\begin{figure}
\begin{center}
\includegraphics[scale=.95]{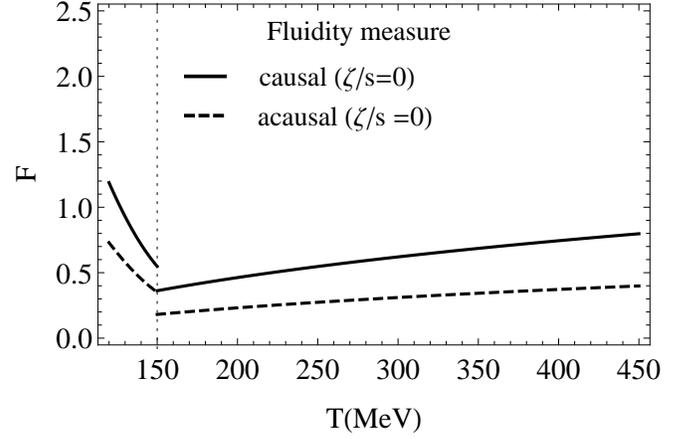}
\caption{Temperature variation of $F$ in the QGP ($T>150$ MeV) 
and hadronic phase ($T<150$ MeV). The verticle line represents $  {\ T_c}=150$ MeV.  } 
\label{f8}
\end{center}
\end{figure}
\subsection{Measure of Fluidity}
First we consider a system devoid of bulk viscosity. Then the fluidity measure of 
such system can be obtained by putting $\zeta=0$ in Eq.\ref{eq46} which leads to:
\begin{equation}
\begin{aligned}
F=\frac{\rho^\frac{1}{3}\Big\{16\eta^2+48\eta^2\beta_2(P+\epsilon)\Big(\frac{\del P}{\del\epsilon}\Big)\Big \}^{1/2}}{\Big[4\Big\{(P+\epsilon)^2\Big(\frac{\del P}{\del \epsilon}\Big)\Big \}^\frac{1}{2} \Big] } 
\label{eq54}
\end{aligned}
\end{equation}
where $\beta_2=3/4P$ in the relativistic limit.
We use  Eq.\ref{eq54} to display the variation of $F$ with 
$T$ for the following inputs.
The particle number density ($\rho$) is estimated from entropy density ($s$) by
using the relation, $\rho\sim s/4$.  
We have used the  parametric form of specific viscosity  given in Ref.~\cite{Hirano} as:
\begin{equation}
\begin{aligned}  
\frac{\eta(T) }{s(T)}&\approx\frac{1}{4\pi }\Big(\frac{s_Q}{s_H}\Big ) \Big (\frac{T}{T_c}\Big ) ^{1-\frac{1}{c^2_s}} \ for\  T<T_c \\& \approx\frac{1}{4\pi }\Big[ 1+W \ln \frac{T}{T_c}\Big ]^2\ for\  T>T_c
\label{eq55}
\end{aligned}
\end{equation}
where $s_Q$ and $s_H$ are the entropy densities in the QGP and hadrons at the 
transition temperature ($T_c=150$ MeV). $W$ is given by
\begin{equation}
\frac{W^2}{4\pi} =\frac{9\beta_0^{'2}}{\Big [80 \pi ^2 K_{SB}\ln\{\frac{4\pi }{g^2(T)}\}\Big ]} 
\label{eq56}
\end{equation} 
where 
\begin{equation}
[g^2(T)]^{-1}=\frac{9}{8}\pi^2\ln (2\pi T/\Lambda)+\frac{4}{9}\pi^2\ln 2[\ln (2\pi T/\Lambda)],
\label{eq57}
\end{equation} 
and $K_{SB}=12$, $\Lambda=190$ MeV  and $\beta_0^\prime=10$. 
The value of entropy density ($s$) and $c_s^2 $ for the hadronic and
QGP phases have been estimated from hadronic resonance gas (HRG) model~\cite{sarwar}
 and quasi-particle QGP model.
The relevant thermodynamic quantities have been derived from the partition
function using standard relations.
The $F$ is displayed as a function of $T$ in Fig.~\ref{f8}
for $\eta/s=1/4\pi$. We observe that
the value of $F$ has increased in the causal scenario compared to the acausal 
dynamics. It is to be also noted that the enhancement is more with larger specific shear viscosity.
The $F$ has a non-linear dependence on the transport coefficients and thermodynamic variables
in causal scenario. However, in the acausal case the dependence on the coefficient of viscosity is linear. This
is reflected in the results already depicted in Fig.~\ref{f8} as well as results displayed below.
We observe a sharp decrease of $F$ in the hadronic phase with the increase in temperature,
{\it i.e.} hadrons flow easily with rise in temperature. However, the temperature variation
of $F$ in QGP phase is slower. As $F$ is larger in causal limit the fluid  flow 
becomes difficult compared to acausal case.  
 
\begin{figure}
\begin{center}
\includegraphics[scale=0.95]{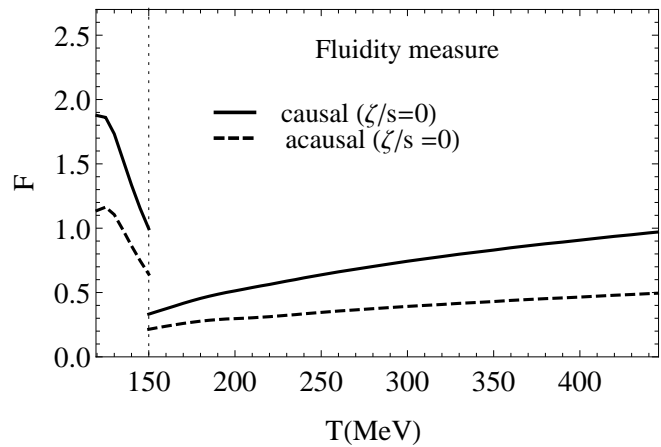}
\caption{Same as Fig.~\ref{f8} with velocity of sound and other
thermodynamic quantities taken from lattice QCD calculations (see text).} 
\label{f9}
\end{center}
\end{figure}

To study the sensitivity of the results on the velocity of sound we use 
the value of  $c_s^2 $ and other relevant thermodynamic variables, like entropy density, etc 
from lattice QCD calculations ~\cite{Katz}.
The variation of $F$ with $T$ is displayed in Fig.~\ref{f9}. A larger discontinuity in $F$
has been seen when $T_c$ and $c_s^2$ are taken from lattice QCD calculations. The shift 
of fluidity  in second order hydrodynamics from the first order is about $35\%$ both in
the hadronic as well as in QGP phase near $T_c$. The same value of $\eta/s$  has been
used for second and first order hydrodynamics, therefore, the shift in $F$ is due to stronger
damping in causal hydrodynamics.

\begin{figure}
\begin{center}
\includegraphics[scale=0.93]{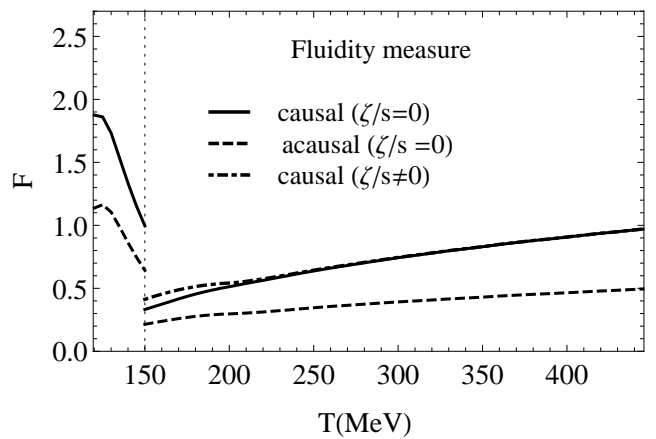}
\caption{Same as Fig.~\ref{f9} in the presence of bulk viscosity (dot-dashed curve).} 
\label{f10}
\end{center}
\end{figure}

\begin{figure}
\begin{center}
\includegraphics[scale=0.95]{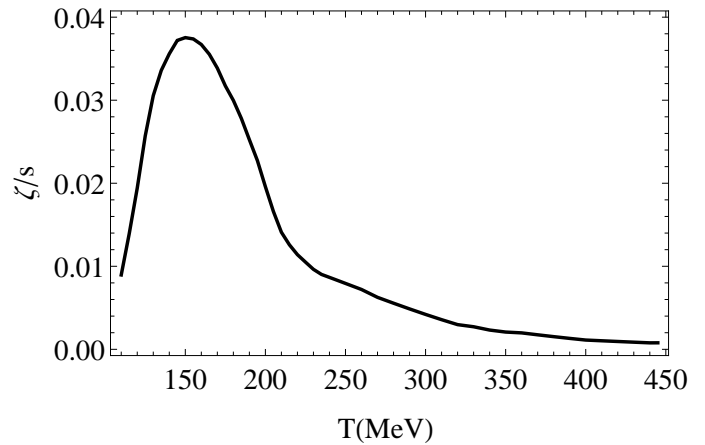}
\caption{Temperature variation of $\zeta /s$.} 
\label{f11}
\end{center}
\end{figure}

\begin{figure}
\begin{center}
\includegraphics[scale=0.95]{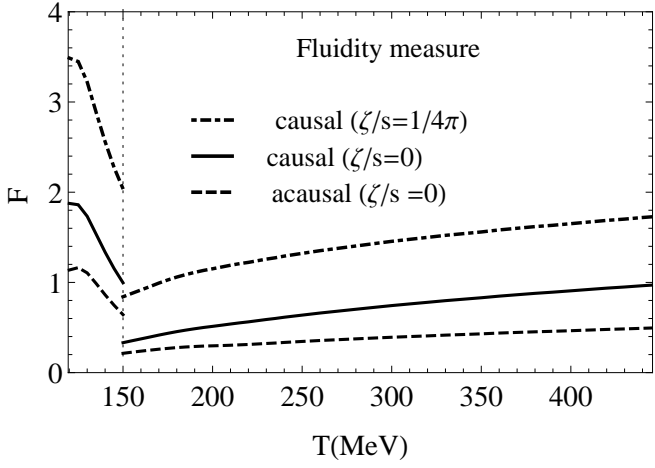}
\caption{Same as Fig.~\ref{f8} in the presence of constant $\zeta/s$ $(=1/4\pi).$ } 
\label{f12}
\end{center}
\end{figure}

Fig.\ref{f10} shows the dependence of fluidity of  QGP on bulk viscosity
in a causal dynamical scenario determined by Eq.\ref{eq46}.     
$\beta _0$, $\beta_2$ and $\beta$ are taken as  $216/P\beta^4$, $3/4P$ and $0.7$ respectively.
The bulk viscosity of the QGP phase has been taken in terms of shear viscosity as~\cite{Weinberg71},
\begin{equation}
\frac{\zeta}{s}\approx 15\frac{\eta }{s}\left(1/3-c_s^2\right)^2  
\label{eq58}
\end{equation}
where the parametric form of $\eta/s$ is taken from Eq.\ref{eq55}. We find 
a peak in the value of $\zeta/s$ around $T\sim 150$ MeV (Fig.~\ref{f11}). This peak is reflected
as a  bump in the temperature variation of $F$ just above $T_c$, 
due to large conformal breaking $({1}/{3}-c_s^2)^2$ near $T_c$. 
It is also interesting to note that the bulk viscosity 
hardly play any role at higher $T$ due to its small numerical value.
As T increases, beyond T=250 MeV, conformal invariance restores and that results 
in almost vanishing $\zeta/s$. 
However, a constant $\zeta/s=1/4\pi$  represents a different picture as 
shown in Fig.~\ref{f12}. It is clear that  non-zero value of $\zeta/s 
(\sim \eta/s)$ will play a crucial role in determining the fluidity
of the system.  

\begin{figure}
\begin{center}
\includegraphics[scale=0.95]{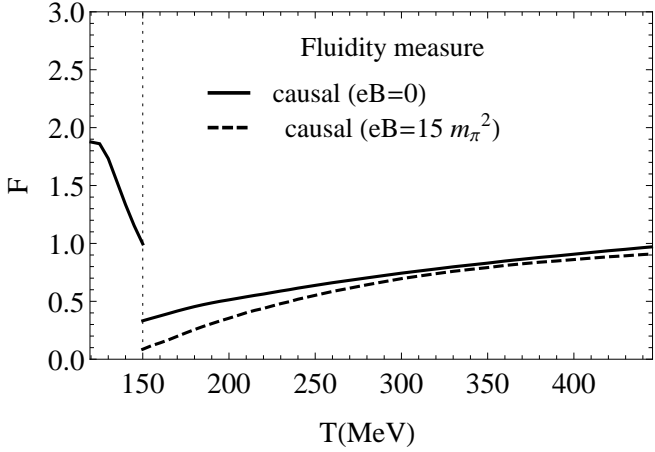}
\caption{Same as Fig.~\ref{f9}
in the presence of magnetic field  ($eB=15m^2_\pi$).} 
\label{f13}
\end{center}
\end{figure}

We have shown before that the magnetic field  alters both the transverse 
and longitudinal modes.
Therefore, it will affect the fluidity of the QGP as shown in Fig.~\ref{f13}. The $F$
for hadronic phase with magnetic field has not been shown, because 
the magnetic field will decay substantially and hence will have insignificant effects on fluidity
of the hadronic phase which appears late in the evolution history. 
As we discussed earlier $B$ makes the fluid less dissipative in the QGP phase. 
Near $T_c$, $F$ reduces significantly and hence the flow becomes easier
near $T_c$. 

\begin{figure}
\begin{center}
\includegraphics[scale=.95]{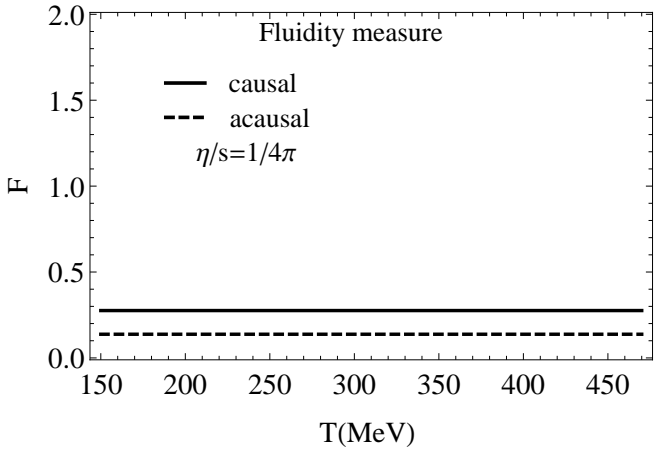}
\caption{Same as Fig.~\ref{f9} for AdS/CFT fluid.} 
\label{f14}
\end{center}
\end{figure}

For AdS/CFT system, we have taken the well known KSS lower bound ($\eta/s=1/4\pi$) of shear 
viscosity~\cite{kss} to show variation of $F$ with $T$ above $T_c$ (Fig.\ref{f14}). 
We have taken  $L_\rho =1/T$ ~\cite{koch_prc} which
gives  $F\approx 0.2$ in acausal hydrodynamics and $F\approx 0.4$ in its causal counterpart.
The fluidity factor $F$ gets enhanced as expected in Israel-Stewart hydrodynamics by 
a factor of 2 hence makes it harder for the fluid to flow.  

\begin{figure}
\begin{center}
\includegraphics[scale=1.]{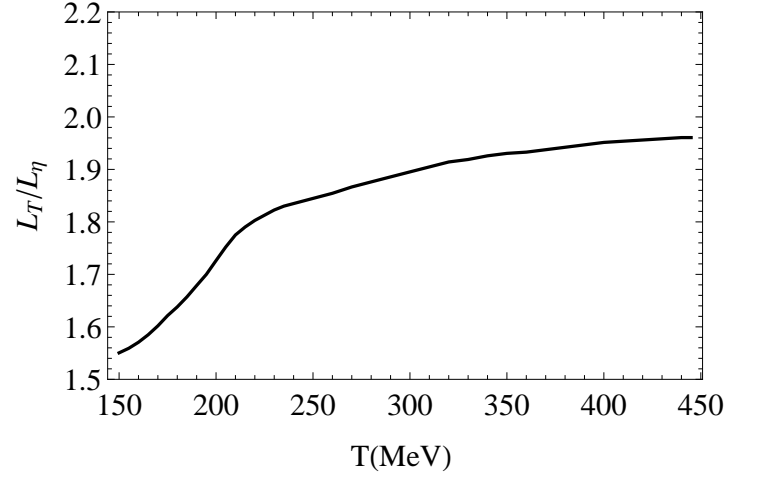}
\caption{The ratio of length scales, $L_T$ and $L_\eta$ corresponding
to causal and acausal hydrodynamics (see text) as a function of temperature.} 
\label{f15}
\end{center}
\end{figure}

In Fig.~\ref{f15} the variation of the ratio of two length scales, $L_T/L_\eta$
has been plotted as a function of $T$. We find that the ratio remains above unity
for the temperature range considered. It is discussed in Ref.~\cite{koch_prc}
that the applicability of hydrodynamics may be resolved from the
the ratio of $L_\eta$ estimated in acausal hydrodynamics to some external
length scale, say, the size of the system, R. Since $L_T/L_\eta>1$, therefore, 
the applicability of hydrodynamics become poorer when causality effects are 
included in the fluid dynamics, if all other relevant quantities kept same
in causal and acausal scenarios.
\section{Summary and Conclusion}
\label{sec:concl}
In summary, we have derived dispersion relations of relativistic fluid using Israel-Stewart second
order causal viscous hydrodynamics. It is shown that the dispersion relations in acausal hydrodynamics 
can be obtained from the causal results as a limiting case. The perturbations in
viscous fluid damp faster within the scope of causal hydrodynamics than its acausal 
counterpart. The waves with large $k$ suffer more  damping than the waves with short  $k$. 
In both the longitudinal and transverse dispersion relations the difference between the causal and 
acausal hydrodynamics is 
significant. The difference increases with the magnitude of viscosities. It has also been
noted that the bulk viscosity does not play any role in the dissipation of the transverse modes
but it plays a crucial role in the dispersion for the longitudinal modes. 
The dispersion relations in the presence of magnetic field 
have also been derived and it is  shown that the magnetic field
affects the longitudinal and transverse modes  oppositely.
The magnetic field makes the fluid effectively less dissipative.
The dispersion relations derived here have been  used to find 
viscous measure of the fluid as well as the viscous horizon. 
We have seen that the use of causal relations enhances the size 
of the viscous horizon of the longitudinal mode by more than a factor two
for the parameter values used here. Inclusion of the causality enhances the $F$ of QGP near $T_c$.
The bulk viscosity affects the fluidity strongly near $T_c$.
However, its role becomes less important at higher temperature with
the restoration of conformal symmetry resulting in lower $\zeta$.  
We also find that the effects of $\zeta$ on $F$ is more prominent 
than $\eta$ if $\eta$ and $\zeta$ have similar magnitudes.
Magnetic field  makes a fluid more perfect by compensating the 
effects of viscosity near $T_c$. The fluidity is enhanced by a constant factor for AdS/CFT fluid  
within the causal hydrodynamics. 

In Ref.~\cite{koch_prc} the fluidity has been studied in the super-critical domain within
the purview of acausal hydrodynamics. We observed a shift in fluidity
due to causal hydrodynamics as compared to acausal hydrodynamics. It is expected that similar
shift will be seen in the super-critical region too.

In a nutshell the incorporation of causality in relativistic hydrodynamics  makes the following
changes with respect to acausal hydrodynamics: (i) The fluidity measure, $F$ increases and thus
flow of the fluid becomes strenuous, (ii) the value  of the highest order of flow harmonics ($n_v$)
reduces as the viscous horizon, $R_v$ increase and (iii) applicability of the hydrodynamics becomes poorer
because $L_T>L_\eta$ in the temperature range considered. In these conclusions it has 
been tacitly assumed that the relevant quantities, like $\eta/s$ etc are kept same in both the
causal and acausal scenarios.

\section{Appendix}
\subsection{Perturbations in  $T^{\lambda\mu }$  }
\label{appa}
We evaluate the perturbation in the energy-momentum tensor $(T^{\lambda\mu })$, of the Israel-Stewart 
hydrodynamics.  We denote perturbations in $P$, $\epsilon$, $n$, $T$ and $u^\alpha$ 
by $P_1,\epsilon_1, n_1, T_1$ and $ u^\alpha_1$ respectively and
decompose $T^{\lambda\mu}$ in Eq.\ref{eq6}  into  sum of
$A^{\lambda\mu}$, $B^{\lambda\mu}$, $C^{\lambda\mu}$, $D^{\lambda\mu}$, $E^{\lambda\mu}$,
and  $F^{\lambda\mu}$.    
We assume the perturbations as $P'=P+P_1$, $\epsilon'=\epsilon+\epsilon_1$, $n'=n+n_1 $, $T'=T+T_1$ 
and $u'=u+u_1$.  
With perturbation $A^{\lambda \mu }(=\epsilon u^\lambda u^\mu+P\triangle^{\lambda\mu})$ changes to 
\begin{equation*}
\begin{aligned}
 A'^{\lambda\mu }&=\epsilon u^\lambda u^\mu +\epsilon_1 u^\lambda u^\mu  +\epsilon  u^\lambda_1 u^\mu+\epsilon  u^\lambda u_1^\mu +P\triangle ^{\lambda\mu}\\&+  P u^\lambda_1 u^\mu+P  u^\lambda u_1^\mu +P_1\triangle ^{\lambda\mu}
\end{aligned}
\end{equation*} 
where we keep only the linear terms in perturbations. Thus, the change in  $A^{\lambda \mu }$  
reads:
 \begin{equation}
 \begin{aligned} A_1^{\lambda\mu }&= \epsilon_1 u^\lambda u^\mu  +\epsilon  u^\lambda_1 u^\mu+\epsilon  u^\lambda u_1^\mu  +  P u^\lambda_1 u^\mu+P  u^\lambda u_1^\mu \\&+P_1\triangle ^{\lambda\mu} 
  \end{aligned}  
\end{equation} 
Similarly the change in the term $B^{\lambda \mu}(=-
\frac{1}{3}\zeta u_{|\sigma} ^\sigma \triangle^{\lambda\mu}+
\frac{1}{9} \zeta^2\beta_0  \dot{u}_{| \rho }^\rho \triangle^ {\lambda\mu})$ arising due from perturbation: 
\begin{equation}
\begin{aligned}
B_1^{\lambda\mu }&=  -\frac{1}{3}\zeta (\del_\sigma u_1^\sigma -\frac{1}{3}\zeta \beta_0\del_0\del_\sigma u_1^\sigma)\triangle^{\lambda\mu } 
\end{aligned}  
\end{equation} 
Perturbation in $C^{\lambda \mu }(=\frac{nT^2}{P+\epsilon}\frac{\zeta \alpha _0\chi  }{3}\del_\sigma [\triangle ^{\sigma\rho }
(\del_\rho \alpha)\triangle^{\lambda\mu}])$ is: 
\begin{align*}C_1^{\lambda\mu }&=\frac{\zeta }{3}\alpha_0\chi \del_\sigma\{(n_1T^2+2TT_1n) \frac{\triangle^{\sigma \rho}}{  P+\epsilon}\del_\rho \alpha \\&-\frac{nT^2 (P_1+\epsilon_1)  }{(P+\epsilon)^2}\triangle^{\sigma \rho}\del_\rho \alpha+\frac{nT^2}{  P+\epsilon}(u^\sigma_1 u^\rho+u^\sigma u^\rho_1) \del_\rho\alpha\}\\&\times \triangle^{\lambda\mu}+\frac{nT^2}{(P+\epsilon)} \frac{\zeta \alpha_0\chi }{3} \del_\sigma \{\triangle^{\sigma\rho}\del_\rho\alpha \}(u^\lambda_1 u^\mu+u^\lambda u^\mu_1)
\end{align*}        
Perturbation in $D^{\lambda\mu}(=-
2\eta u^{<\lambda|\mu >})$ reads: 
\begin{equation}
\begin{aligned}D_1^{\lambda\mu }= -\eta(\triangle _\alpha ^\lambda\triangle_\beta ^\mu +\triangle _\alpha ^\mu\triangle_\beta ^\lambda-\frac{2}{3}\triangle ^{\lambda\mu }\triangle _{\alpha\beta })\del ^\beta u^\alpha _1
\end{aligned}        
\end{equation} 
Change in the term $E^{\lambda\mu}(=4\eta^2\beta_2\dot{u}^{<\lambda|\mu >})$ due to perturbation is:
         \begin{equation}
         \begin{aligned}E_1^{\lambda\mu }= 2\eta^2\beta_2\del _0\{(\triangle _\alpha ^\lambda\triangle_\beta ^\mu +\triangle _\alpha ^\mu\triangle_\beta ^\lambda-\frac{2}{3}\triangle ^{\lambda\mu }\triangle _{\alpha\beta })\del ^\beta u^\alpha _1\}
          \end{aligned}        
        \end{equation}
The $F^{\lambda\mu }(=\frac{2nT^2}{P+\epsilon}\alpha _1\eta\chi [\triangle^{(\lambda} _\alpha\triangle^{\mu)} _{\beta}-\frac{1}{3}\triangle_{\alpha \beta}
\triangle^{\lambda\mu }] \del^{\beta} \triangle^{\alpha \rho }\del_\rho \alpha )$ is perturbed by the term:
\begin{equation}
\begin{aligned}
& F_1^{\lambda\mu } = \alpha_1\chi \eta [\triangle _\alpha ^\lambda\triangle_\beta ^\mu +\triangle _\alpha ^\mu\triangle_\beta ^\lambda- \frac{2}{3}\triangle ^{\lambda\mu }\triangle _{\alpha\beta }]\del^\beta\\&\times [\{\frac{n_1T^2+2TT_1 n}{P_1+\epsilon_1}- \frac{n T^2(P_1+\epsilon_1)}{(P +\epsilon )^2} \}\triangle ^{\alpha \rho}\del _\rho \alpha \\& +\frac{nT^2}{P+\epsilon}(u^\rho_1 u^\alpha  +u^\rho u^\alpha_1)\del_\rho\alpha]   +\frac{nT^2}{P+\epsilon}\alpha_1\chi \eta[(u_1^\lambda u_\alpha \\&  +u ^\lambda u_{ 1 \alpha}) \triangle ^\mu_\beta  + (u_1^\mu u_\beta  +u ^\mu u_{ 1 \beta}) \triangle ^\lambda_\alpha+(u_1^\mu u_\alpha\\& +u ^\mu u_{ 1 \alpha}) \triangle ^\lambda_\beta   + (u_1^\lambda u_\beta  +u ^\lambda u_{ 1 \beta}) \triangle ^\mu_\alpha      -\frac{2}{3}\{(u_1^\lambda u_\mu\\& +u ^\lambda u_{ 1 \mu})\triangle _{\alpha \beta} +(u_{1\alpha} u_\beta     +u_\alpha u_{1\beta })   \triangle ^{\lambda \mu }\}]     \del ^\beta [\triangle^{\alpha \rho} \del_\rho \alpha]
 \end{aligned}
 \end{equation}
 The net change in $T^{\lambda\mu }$ due to perturbation  is the sum of all terms discussed above: 
   \begin{equation}
   T_1^{\lambda\mu }=A_1^{\lambda\mu }+B_1^{\lambda\mu }+C_1^{\lambda\mu } +D_1^{\lambda\mu }+E_1^{\lambda\mu }+F_1^{\lambda\mu }
   \end{equation}
\subsection{Disperson relation for the longitudinal mode.}
\label{appb}
The linearized equation of motion (EoM) of the Israel-Stewart hydrodynamics can be written in terms of the independent variables
(perturbations)
{\it e.g.} $(\vec{k}\cdot\vec{u_1}),  T_1$ and $n_1$. Then the  dispersion relation can be obtained  by setting the 
determinant of the coefficients of the linear algebraic equations satisfied by
$(\vec{k}\cdot\vec{u_1}), T_1$ and $n_1$ to zero.  
Expanding this determinant and solving for $\omega$ leads to the dispersion relation for longitudinal component.
The determinant formed by three unknown coefficients in Eqs.\ref{eq10}, \ref{eq11} and \ref{eq12} is:
 \begin{align*}
 0=\begin{vmatrix}
   a_{11} & a_{12} & a_{13}\\  
   a_{21} & a_{22} & a_{23} \\
   a_{31} & a_{32} & a_{33} \\
 \end{vmatrix}
 \end{align*}
where the values of the different matrix elements are given below:
  \begin{equation*}
  \begin{aligned}
   a_{11}& =\omega (P+\epsilon)+\frac{nT^2}{ 3(\epsilon+P)}\zeta\chi \alpha_0\omega \nabla^2\alpha+ \frac{2 nT^2}{(\epsilon+P)} \alpha_1\eta\omega \chi\\& \times \{\frac{(\vec{k}\cdot\vec{\nabla})(\vec{k}\cdot\vec{\nabla})\alpha}{k^2}-\frac{1}{3}\nabla^2\alpha\}+i\frac{\zeta}{3} k^2 -\frac{1}{9}\zeta^2 \beta_0\omega k^2\\&-\frac{nT^2}{3(P+\epsilon)}\zeta \alpha_0\chi[2(\vec{k}\cdot\vec{\nabla})\dot{\alpha }+i\dot\alpha k^2-i\omega (\vec{k}\cdot\vec{\nabla})\alpha]\\&+i\frac{4}{3}\eta k^2-\frac{8}{3}\eta^2\beta_2\omega k^2-\frac{8nT^2}{3(P+\epsilon)}\alpha_1\chi\eta(\vec{k}\cdot\vec{\nabla})\dot{\alpha} \\&-i \frac{4 \alpha_1\eta\chi}{3}  k^2 \dot{\alpha }
\end{aligned}
\end{equation*}
\begin{equation*}
\begin{aligned}
a_{12}&=-k^2\Big(\frac{\del P}{\del T}\Big)_n-\Re \Big[\frac{ 2nT }{P+\epsilon}-\frac{ nT^2}{(P+\epsilon)^2}\{\Big(\frac{\del P}{\del T}\Big)_n\\& +\Big(\frac{\del \epsilon}{\del T}\Big)_n\}\Big]
\end{aligned}
\end{equation*}
\begin{equation*}
\begin{aligned}
a_{13} &= -k^2\Big(\frac{\del P}{\del n}\Big)_T-\Re \Big[\frac{ T^2}{P+\epsilon}-\frac{ nT^2}{(P+\epsilon)^2}\{\Big(\frac{\del P}{\del n}\Big)_T\\& +\Big(\frac{\del \epsilon}{\del n}\Big)_T\}\Big] 
\end{aligned}
\end{equation*}
\begin{equation*}
\begin{aligned}
a_{21}&=-(\epsilon +P)-\frac{\zeta}{3} \frac{nT^2}{(\epsilon+P)}\chi\alpha_0\nabla^2\alpha-\frac{2nT^2}{(\epsilon+P)}\alpha _1\eta \chi\\&\times \{\frac{(\vec{k}\cdot\vec{\nabla})(\vec{k}\cdot\vec{\nabla})\alpha}{k^2}- \frac{1}{3}\nabla^2\alpha\} 
\\& a_{22}=\omega \Big(\frac{\del 
\epsilon}{\del T}\Big)_n 
\\& a_{23}=\omega \Big(\frac{\del \epsilon}{\del n}\Big)_T 
\\& a_{31}=-n
\\& a_{32}=0
\\& a_{33}=\omega
\end{aligned}
\end{equation*}
where
  \begin{equation*}
  \begin{aligned}
   \Re &\equiv i\frac{\zeta \alpha_0\chi}{3}    k^2(\vec{k}\cdot\vec{\nabla})\alpha+\frac{\zeta \chi\alpha_0}{3}k^2\nabla^2\alpha \\& +i \frac{4 \alpha_1\eta\chi}{3}  k^2(\vec{k}\cdot\vec{\nabla})\alpha
  \end{aligned}
  \end{equation*}
Expanding the above determinant and keeping terms  upto 2nd order 
in $\eta T/h, \zeta T/h$ and their products like 
$\eta \chi ,\eta \zeta ,\zeta \chi$, we get an equation of the form
  \begin{equation}
  \omega (a \omega^2 +b\omega +c )=0
  \label{gendisp}
  \end{equation} 
which has a trivial solution $\omega =0$ and the other two roots can be found
by solving the quadratic equation $(a \omega^2 +b\omega +c )=0$. 
The coefficients of the quadratic equation is given by
\begin{equation}
\begin{aligned}
  a &=  \Big[(P+\epsilon)+\frac{nT^2}{3(P+\epsilon)}\zeta\chi \alpha _0\nabla^2\alpha -\frac{1}{9}k^2\zeta^2\beta_0      \\& +  \frac{ 2nT^2\eta\chi\alpha_1}{(P+\epsilon)}\frac{(\vec{k}\cdot\vec{\nabla})(\vec{k}\cdot\vec{\nabla})\alpha}{k^2} -\frac{2nT^2}{3(P+\epsilon)}\eta\chi \alpha _1\nabla^2\alpha     \\& -\frac{8}{3}k^2\eta^2\beta_2       +i\frac{nT^2}{3(P+\epsilon)}\zeta\chi \alpha _0 (\vec{k}\cdot\vec{\nabla})\alpha \Big] \Big (\frac{\del \epsilon}{\del T}\Big)_n
  \end{aligned}
  \end{equation}
  
  \begin{equation}
   \begin{aligned}
   b &=\Big[-\frac{2}{3}\frac{ nT^2}{(P+\epsilon)} \zeta\chi\alpha_0  (\vec{k}\cdot\vec{\nabla})\dot{\alpha} -\frac{8}{3}\frac{ nT^2\eta\chi\alpha_1}{(P+\epsilon)}(\vec{k}\cdot\vec{\nabla})\dot{\alpha}\\&  +i\Big\{\frac{1}{3}k^2\zeta +\frac{4}{3}k^2\eta   -\frac{1}{3}\frac{ nT^2}{(P+\epsilon)} k^2\dot{\alpha}\zeta\chi\alpha_0 \\& -\frac{4}{3}k^2\alpha_1\eta\chi\dot{\alpha} \Big\}\Big]\Big (\frac{\del \epsilon}{\del T}\Big)_n
   \end{aligned}
   \end{equation}
   \begin{equation}
   \begin{aligned}
   c&=  - k^2(P+\epsilon)\Big(\frac{\del P}{\del T}\Big)_n+k^2n\Big(\frac{\del P}{\del T}\Big)_n\Big(\frac{\del \epsilon}{\del n}\Big)_T\\&-k^2n\Big(\frac{\del P}{\del n}\Big)_T
   \Big(\frac{\del \epsilon}{\del T}\Big)_n  -\frac{2}{3} k^2nT\zeta\chi\alpha_0\nabla^2\alpha\\&+\frac{2n^2T}{(P+\epsilon)}\alpha_0\zeta\chi k^2\Big(\frac{\del \epsilon}{\del n}\Big)_T \nabla^2\alpha  \\&-\frac{n^2T^2}{3(P+\epsilon)^2 }\alpha_0\zeta\chi k^2\Big(\frac{\del \epsilon}{\del n}\Big)_T\Big(\frac{\del P}{\del T}\Big)_n\nabla^2\alpha  \\& +\frac{n^2T^2}{3(P+\epsilon)^2}\alpha_0\zeta\chi k^2\Big(\frac{\del \epsilon}{\del T}\Big)_n\Big(\frac{\del P}{\del n}\Big)_T\nabla^2\alpha \\& +\frac{2n  T^2 }{3(P+\epsilon) }  \alpha_1\eta\chi k^2  \Big(\frac{\del P}{\del T}\Big)_n\nabla^2\alpha    -\frac{2nT^2}{P+\epsilon}\eta\chi\alpha_1\Big(\frac{\del P}{\del T}\Big)_n  \\&\times (\vec{k}\cdot\vec{\nabla})(\vec{k}\cdot\vec{\nabla})\alpha   +i\Big[-\frac{2}{3}k^2nT\zeta\chi\alpha_0\\& +\frac{nT^2}{P+\epsilon}\zeta\chi\alpha_0 k^2\Big(\frac{\del P}{\del T}\Big)_n +\frac{2n^2T }{3(P+\epsilon)}\zeta\chi\alpha_0 k^2\\&- \frac{ n^2T^2 }{3(P+\epsilon)^2}\zeta\chi\alpha_0 k^2\Big(\frac{\del P}{\del T}\Big)_n\Big(\frac{\del \epsilon}{\del n}\Big)_T\\& +\frac{ n^2T^2 }{3(P+\epsilon)^2}\zeta\chi\alpha_0 k^2\Big(\frac{\del P}{\del n}\Big)_T\Big(\frac{\del \epsilon}{\del T}\Big)_n  -\frac{8}{3}k^2nT\eta\chi\alpha_1\\&+\frac{ 4nT^2 }{3(P+\epsilon)}\eta\chi\alpha_1 k^2\Big(\frac{\del P}{\del T}\Big)_n+\frac{ 8n^2T }{3(P+\epsilon)}\eta\chi\alpha_1 k^2\Big(\frac{\del \epsilon }{\del n}\Big)_T\\&-\frac{ 4n^2T^2 }{3(P+\epsilon)^2}\eta\chi\alpha_1 k^2  \Big(\frac{\del \epsilon }{\del n}\Big)_T\Big(\frac{\del P }{\del T}\Big)_n\\&+\frac{ 4n^2T^2 }{3(P+\epsilon)^2}\eta\chi\alpha_1 k^2\Big(\frac{\del \epsilon }{\del T}\Big)_n\Big(\frac{\del P }{\del n}\Big)_T\Big](\vec{k}\cdot\vec{\nabla})\alpha
   \end{aligned}
   \end{equation}
   
The physical solution of Eq.\ref{gendisp} gives the general dispersion relation for 
 non-zero $\eta,\zeta,\chi$ as well for non-zero (baryonic) conserved charge.

 \subsection{Perturbations in  $T^{\lambda\mu }$ in the presence of magnetic field}
 \label{appc}
The energy-momentum tensor in the presence of magnetic field$(B)$ is given by Eq.\ref{eq23} with 
$n'^\mu={B^\mu}/{B}$ where $B_\mu =(1/2)\epsilon_{\mu\nu\alpha\beta}F^{\nu\alpha}u^\beta$ and 
$F^{\mu\nu }=(E^\mu u^\nu -E^\nu u^\mu )+
(1/2)\epsilon^{\mu\nu\beta\gamma}(u_\beta B_\gamma-u_\gamma B_\beta)$. 
For vanishing electric field $(E)$, the expression for energy-momentum
tensor (Eq.\ref{eq23}) can be written as:
 \begin{equation}
 \begin{aligned}
  T^{\lambda \mu }_m=\frac{B^2}{8\pi }(2u^\lambda u^\mu +g^{\lambda\mu}-\frac{1}{8}\epsilon^{\lambda\rho \sigma\eta}\epsilon_{\rho\sigma\alpha\beta}\epsilon^{\rho'\sigma'\eta'\mu}\epsilon_{\rho'\sigma'\alpha'\beta'} &\\  (u^\alpha B^\beta -u^\beta  B^\alpha)(u^{\alpha'} B^{\beta'}-u^{\beta'} B^{\alpha'} )u_\eta u_{\eta'}
  \label{eqc1}
  \end{aligned}
 \end{equation}
 Using the following relation satisfied by the Levi-civita tensor
 \begin{equation*}
 \epsilon^{\rho \sigma\eta\lambda}\epsilon_{\rho\sigma\alpha\beta}=-2(g^\eta _\alpha g^\lambda _\beta -g^\eta _\beta g^\lambda _\alpha) 
 \end{equation*} 
Eq.\ref{eqc1} can be written as:
 \begin{equation}
 \begin{aligned}
  T^{\lambda \mu }_m=
  \frac{B^2}{8\pi } [2u^\lambda u^\mu 
  +g^{\lambda\mu}-
  \frac{1}{2B^2}(g^\eta 
  _\alpha g^\lambda _\beta -g^\eta _\beta g^\lambda _\alpha)&\\(g^{\eta'} _{\alpha'}  g_{\beta'}^\mu  -g^{\eta'}_{\beta'} g^\mu _{\alpha'}) 
      (u^\alpha B^\beta u^{\alpha'} 
     B^{\beta'}
     -u^\alpha B^{\beta}
      u^{\alpha'} B^{\beta'}
       &\\-u^\beta  B^\alpha u^{\alpha'}
       B^{\beta'} +u^\beta  B^\alpha u^{\beta'}
      B^{\alpha'})u_\eta u_{\eta'}]
      \label{c2}
  \end{aligned}
 \end{equation}
The last term of the RHS of Eq.\ref{c2} can be decomposed into 16 terms, 
each of  which will contain product of four fluid velocity ($u^\mu$). 
If we write, $u^\mu\rightarrow u^\mu+u_1^\mu$ and use $u^0=1$, $u^i=0$,
$u_1^0=0$, $B^0=0$ and $B^\mu u_\mu=0$, then the only non-zero
terms are: $(1/2)u^\alpha u^{\alpha '}u_{1\eta }u_{\eta '}B^\beta B^{\beta'}g^\eta _\beta g^\lambda _\alpha g^{\eta'}_{\alpha '}g^\mu _{\beta'}$ 
 and $(1/2)u^\beta u^{\beta'}u_{1\eta }u_{\eta'}B^\alpha B^{\alpha'}g^\eta _\alpha g^\lambda _\beta g^{\eta'}_{\beta'}g^\mu _{\alpha'}$. 
From these two terms the perturbations are estimated. . 
The only non-zero components of the perturbation, 
$T^{\lambda\mu}_{1m}$ to  $T^{\lambda\mu}_m$ exist for $\lambda=i$ and $\mu=0$.
The magnitude of the perturbations are:
  \begin{equation}
  \begin{aligned}
  \\& T^{00}_{1m} =0
     \\& T^{i0}_{1m} =\frac{B^2}{8\pi }(2u^i_1-\frac{B^i}{B^2}(\vec{u_1}\cdot \vec{B}))
     \\& T^{ij}_{1m}=0.
  \label{eqc3}
  \end{aligned}  
  \end{equation}

These expressions for the energy momentum tensor due to presence of magnetic field in the fluid
have been used to calculate the dispersion relation for the longitudinal and
transverse wave in this work.

{\bf Acknowledgment :}
We are grateful to Golam Sarwar for many helpful discussions and MR would like to thank
Department of Atomic Energy, Govt of India for financial support. We  also  thankful to S. D.  Katz for providing the required 
lattice data.

\end{document}